\documentclass[12pt,reqno]{amsart}
\usepackage[dvips,colorlinks=true,linkcolor=red,citecolor=blue]{hyperref}
\usepackage[final]{graphicx}
\usepackage{amsfonts}
\usepackage{pdfsync}
\usepackage{amsmath}
\usepackage{amssymb}
\usepackage{amsthm}
\newtheorem{theorem}{Theorem}[section]
\newtheorem{lemma}[theorem]{Lemma}
\newtheorem{corollary}[theorem]{Corollary}
\newtheorem{proposition}[theorem]{Proposition}

\newcommand{\R}{{\mathord{\mathbb R}}}
\newcommand{\C}{{\mathord{\mathbb C}}}

\newcommand{\Z}{{\mathord{\mathbb Z}}}
\newcommand{\N}{{\mathord{\mathbb N}}}
\newcommand{\E}{{\mathord{\mathbb E}}}
\newcommand{\PP}{{\mathord{\mathbb P}}}
\newcommand{\supp}{{\mathop{\rm supp\ }}}
\newcommand{\tr}{{\mathop{\rm tr\ }}}

\newcommand{\esp}{\mathbb{E}}
\newcommand{\pro}{\mathbb{P}}

\begin{document}

\title[The random displacement model]{Understanding the Random
  Displacement Model:\\ From Ground-State Properties to Localization}

\author[Klopp]{Fr{\'e}d{\'e}ric Klopp$^1$}

\thanks{F.\ K.\ was partially supported by the ANR grant
  08-BLAN-0261-01.}

\author[Loss]{Michael Loss$^2$}

\thanks{M.\ L.\ was supported in part by NSF grant DMS-0901304.}

\author[Nakamura]{Shu Nakamura$^3$}

\thanks{S.\ N.\ was partially supported by JSPS grant Kiban (A)
  21244008}

\author[Stolz]{G{\"u}nter Stolz$^4$}

\thanks{G.\ S.\ was supported in part by NSF grant DMS-0653374.}

\address{$^1$ LAGA, U.M.R. 7539 C.N.R.S, Institut Galil{\'e}e, Universit{\'e}
  de Paris-Nord, 99 Avenue J.-B.  Cl{\'e}ment, F-93430 Villetaneuse,
  France\ et \ Institut Universitaire de France}
\email{\href{mailto:klopp@math.univ-paris13.fr}{klopp@math.univ-paris13.fr}}

\address{$^2$ Georgia Institute of Technology, School of Mathematics,
  Atlanta, Georgia 30332-0160, loss@math.gatech.edu}
\email{\href{mailto:loss@math.gatech.edu}{loss@math.gatech.edu}}

\address{$^3$ Graduate School of Mathematical Sciences, University of
  Tokyo, 3-8-1 Komaba, Meguro-ku, Tokya, Japan 153-8914,
  shu@ms.u-tokyo.ac.jp}
\email{\href{mailto:shu@ms.u-tokyo.ac.jp}{shu@ms.u-tokyo.ac.jp}}

\address{$^4$ University of Alabama at Birmingham, Department of
  Mathematics, Birmingham, Alabama 35294-1170, stolz@uab.edu}
\email{\href{mailto:stolz@uab.edu}{stolz@uab.edu}}

% \date{\today}
\maketitle

\vspace{.3truein} \centerline{\bf Abstract}

\medskip {\sl We give a detailed survey of results obtained in the
  most recent half decade which led to a deeper understanding of the
  random displacement model, a model of a random Schr{\"o}dinger operator
  which describes the quantum mechanics of an electron in a
  structurally disordered medium. These results started by identifying
  configurations which characterize minimal energy, then led to
  Lifshitz tail bounds on the integrated density of states as well as
  a Wegner estimate near the spectral minimum, which ultimately
  resulted in a proof of spectral and dynamical localization at low
  energy for the multi-dimensional random displacement model.}

\section{Introduction}
\label{sec:introduction}

By the random displacement model (RDM) we refer to a random
Schr{\"o}dinger operator of the type
\begin{equation} \label{eq:RDM} H_{\omega} = -\Delta + V_{\omega},
  \quad V_{\omega}(x):=\sum_{n\in \Z^d} q(x-n-\omega_n)
\end{equation}
in $L^2(\R^d)$, $d\ge 1$. The potential is generated by randomly
displacing translates of the single-site potential $q$ from the
lattice sites $n\in \Z^d$. More detailed assumptions on $q$ and the
random displacements $\omega_n$ will be introduced below as needed.

The RDM has proven to be much harder to analyze mathematically than
the (continuum) Anderson model
\begin{equation}
  \label{eq:Anderson} H_{\lambda(\omega)}^A=-\Delta + \sum_{n\in
    \Z^d} \lambda_n q(x-n)
\end{equation}
with random coupling constants $\lambda_n=\lambda_n(\omega)$. A
fundamental technical difference between the RDM and the Anderson
model lies in their monotonicity properties. If the single-site
potential $q$ is sign-definite, then the Anderson model is monotone in
the random variables $\lambda_n$ in quadratic form sense. This is not
true for the RDM, independent of sign-assumptions on $q$.

Many of the rigorous tools which have been developed to study the
Anderson model rely on its monotonicity properties. In particular,
this is true for most of the proofs of localization for the Anderson
model near the bottom of its spectrum. In fact, if one considers the
Anderson model with sign-indefinite single-site potential $q$, and
thus looses monotonicity, then localization results are much more
recent and far less complete than for the case of sign-definite $q$,
e.g.\ \cite{Veselic, Klopp95, HislopKlopp, KN1, KN2}. The difficulties which
arise are in many ways similar to the problems encountered in the RDM. Related phenomena and difficulties also arise in discrete alloy-type models with sign-indefinite single site potential, as recently reviewed in \cite{EKTV}.

Among models for continuum random Schr{\"o}dinger operators, the {\it
  structural disorder} described by the RDM can be considered as
physically equally natural as the coupling constant disorder in the
Anderson model. Another natural model for structural disorder is the
Poisson model
\begin{equation} \label{eq:Poisson} -\Delta + \sum_i q(x-X_i),
\end{equation}
with $X_i$ denoting the points of a $d$-dimensional Poisson
process. The RDM as well as the Poisson model were introduced early on
in the mathematical literature on continuum random Schr{\"o}dinger
operators, e.g.\ \cite{Kirsch, PasturFigotin} and references
therein. However, progress has been much more limited than for the
Anderson model due to the technical difficulties which arise.

An exception is the case $d=1$, where localization throughout the
entire spectrum has been proven for the RDM and the Poisson model in
\cite{Stolz, BS, DSS}. This was possible based on the powerful
dynamical systems methods available to study one-dimensional random
operators, in particular those allowing to prove positivity of
Lyapunov exponents and to deduce localization from this. However, the
non-monotonicity of the RDM and the Poisson model has visible
consequences already in the one-dimensional case, for example through
the appearance of critical energies in the spectrum at which the
Lyapunov exponent vanishes and, in some cases, weaker results (e.g.\
on dynamical localization, which has not been shown for the
one-dimensional Poisson model).

In dimension $d\ge 2$ it is generally expected that ``typical'' random
Schr{\"o}dinger operators have a localized region at the bottom of the
spectrum, at least if the latter corresponds to a {\it fluctuation
  boundary} of the spectrum, which describes a boundary characterized
by rare events. The history of localization proofs for the
multi-dimensional RDM and Poisson model is told very quickly. For the
Poisson model in $d\ge 2$ localization at the bottom of the spectrum
was finally proven in \cite{GHK1} for positive single-site potentials
and in \cite{GHK2} for negative single-site potentials. In both cases
the powerful extension of multi-scale analysis developed by Bourgain
and Kenig in \cite{BK} was used as a tool.

There were two previous results on localization for the
multi-dimensional RDM, \cite{Klopp} and \cite{Ghribi/Klopp}. In
\cite{Klopp} a semiclassical version of (\ref{eq:RDM}) is considered
and localization near the bottom of the spectrum is established for
sufficiently small values of a semiclassical coupling parameter at the
Laplacian. \cite{Ghribi/Klopp} considers the RDM with an additional
periodic term $V_{per}$ and establishes localization for generic (but
non-zero) choices of $V_{per}$. In both works the values of the
displacements $\omega_n$ have to be sufficiently small and first order
perturbation effects (such as a monotonicity property of Floquet
eigenvalues of $-\Delta+V_{per}$ in \cite{Ghribi/Klopp}) are
exploited. What makes the ``naked RDM'' (\ref{eq:RDM}) more difficult
to handle is that, as will be pointed out below, one ultimately has to
resort to second-order perturbation effects.

The goal of this work is to give a detailed survey of new results for
the RDM (\ref{eq:RDM}) obtained in the papers \cite{BLS1, BLS2, KN2}
and \cite{KLNS}, which allowed to understand that the spectral minimum
of the RDM is a fluctuation boundary under a natural set of
assumptions not requiring additional parameters or smallness of the
displacements (other than a non-overlap condition), and ultimately led
to a proof of localization in this setting in \cite{KLNS}.

The strategy used to prove localization in these works is the one
provided by the Fr{\"o}hlich-Spencer multi-scale analysis \cite{FrSp}, as
described for continuum models in very accessible form in the book
\cite{Stollmann}, and with state-of-the-art results shown in
\cite{Germinet/Klein} and surveyed in \cite{Klein}. In essence, the
MSA approach shows that localization, spectral as well as dynamical,
can be proven once a smallness result (``Lifshitz tails'') for the
integrated density of states at the bottom of the spectrum and a
Wegner estimate are available as input.

Therefore much of our effort is aimed at proving these two
ingredients. However, for the RDM (\ref{eq:RDM}) one first needs to
address a preliminary problem: Which configurations
$\omega=(\omega_n)$ characterize the minimum of the almost sure
spectrum of $H_{\omega}$? To explain that this is a non-trivial issue,
let us compare with the Anderson and Poisson models. In the Anderson
model (\ref{eq:Anderson}), due to monotonicity, the spectrum is
minimized by choosing all coupling constants $\lambda_n$ minimal (in
the support of their distribution) if $q$ is positive, while all
$\lambda_n$ should be chosen maximal if $q$ is negative. For the
Poisson model (\ref{eq:Poisson}) the spectral minimum is $0$ if $q$ is
positive, corresponding to regions with widely separated Poisson
points. If $q$ is negative, then regions of densely clustered Poisson
points lead to spectral minimum $-\infty$. The mechanism for
generating the spectral minimum in the RDM is much less apparent (with
similar difficulties arising for the Anderson model with
non sign-definite single-site potential). In fact, while for the
(definite) Anderson and Poisson model the spectrum is minimized by
minimizing the potential, for the RDM we will see that a much more
subtle interaction between kinetic and potential energy determines the
spectral minimum.

In terms of assumptions to be made, the most important one is that the
single-site potential shares the symmetries of the underlying lattice,
here $\Z^d$. It is fair to say that in our approach symmetry replaces
the lack of any apparent monotonicity properties of the model,
ultimately allowing to identify more delicate monotonicity properties
which are at the core of our proofs of Lifshitz tail bounds and a
Wegner estimate for the RDM.

We find it remarkable how many mathematical ideas and tools had to be
invoked and how all this ultimately fit together quite perfectly to
lead to a localization proof for the RDM (\ref{eq:RDM}). Getting this
across to the reader is our main motivation for providing this
expository account of our work. Beyond merely stating a series of
results, we include frequent discussions of the underlying
motivations, often going beyond what we have been able to include in
our previously published work. We have also tried to include at least
outlines of all proofs, even if we frequently have to refer to the
original papers for additional details.

A rough outline of the contents of the remaining sections of this
paper is as follows: In Section~\ref{sec:minimum} we reveal how the
spectral minimum of the RDM is found. In Section~\ref{sec:Neumann} we
show how the proof of this is reduced to a spectral minimization
property of a related single-site Neumann operator. This operator and
its ground state properties are central to almost all our results. In
particular, we will revisit this operator in
Section~\ref{sec:missinglink} and explain why we ultimately needed to
know more about it than what is stated in
Section~\ref{sec:Neumann}. To avoid having to interrupt the telling of
our localization story, we outline the proofs of these results in
Section~\ref{sec:bubblesresults} near the end of the paper.

The rest of the localization story is told in
Sections~\ref{sec:uniqueness}, \ref{sec:SpecLifshitz},
\ref{sec:GenLifshitz}, \ref{sec:Wegner} and
\ref{sec:localization}. Section~\ref{sec:uniqueness} yields
information on uniqueness of configurations characterizing the
spectral minimum which is necessary for the proof of Lifshitz tail
bounds in Sections~\ref{sec:SpecLifshitz} and
\ref{sec:GenLifshitz}. The results in the latter two sections work in
form of a boot-strap, starting with a Lifshitz tail bound under
strong additional assumptions which are then relaxed. Our Wegner
estimate for the RDM is presented in Section~\ref{sec:Wegner}. In
Section~\ref{sec:localization} we state the exact form of our result
on localization for the RDM and provide references to the literature
on multi-scale analysis, which show how this is proven based on the
Lifshitz tail and Wegner bounds. In the very last
Section~\ref{sec:problems} we discuss some open problems related to
our work.

\vspace{.3cm}

\noindent {\bf Acknowledgments:} G.\ S.\ would like to thank the
organizers of the conference ``Spectral Days'', held in Santiago from
September 20 to 24, 2010, for the opportunity to give a lecture series
on the work discussed here.

\section{The Spectral Minimum of the RDM} \label{sec:minimum}

We will always assume that the displacement parameters $\omega =
(\omega_n)_{n\in \Z^d}$ are independent, identically distributed
$\R^d$-valued random variables. Their common distribution is a Borel
probability measure on $\R^d$. As usual, we define its support by
\[ \mbox{supp}\, \mu := \{ a\in \R^d: \mu(B_{\varepsilon}(a))>0
\;\mbox{for all}\; \varepsilon>0\},\] which is a closed set. The
i.i.d.\ random variables $\omega_n$ can be realized as the canonical
projections $\omega \mapsto \omega_n$ in the infinite product
probability space
\[ (\Omega, \PP) = \left( \otimes_{n\in \Z^d} \R^d, \otimes_{n\in
    \Z^d} \mu \right).\]

Under weak assumptions on $\mu$ and the single-site potential $q$ the
RDM $H_{\omega}$ is self-adjoint on the second order Sobolev space in
$L^2(\R^d)$ and ergodic with respect to shifts in $\Z^d$ in the sense
of, e.g., \cite{Carmona/Lacroix}. Thus its spectrum is almost surely
deterministic: There exists $\Sigma \subset \R$ such that
\begin{equation} \label{eq:asspectrum} \sigma(H_{\omega}) = \Sigma
  \quad \mbox{for $\PP$-almost every $\omega$}.
\end{equation}

In fact, one has
\begin{equation} \label{eq:persupp} \Sigma = \overline{\bigcup_{\omega
      \in {\mathcal C}_{per}} \sigma(H_{\omega})},
\end{equation}
where
\begin{equation} \label{eq:perconfig} {\mathcal C}_{per} := \{ \omega
  \in \Z^d \to \;\mbox{supp}\,\mu \:\mbox{periodic with respect to a
    sub-lattice of}\: \Z^d\}.
\end{equation}
This follows by the same methods which have been used to prove a corresponding result
for the Anderson model: That $\Sigma$ is contained in the right hand side of (\ref{eq:persupp}) follows by approximating any given random configuration with periodic configurations, truncating the random configuration to large cubes and periodically extending from there. On the other hand, one can show that almost every random configuration comes arbitrarily close to any given periodic configuration on arbitrarily large cubes, which is he idea behind the reverse inclusion. For a detailed proof, written for the case of the Anderson model, we refer to \cite{Kirsch}.

In particular, (\ref{eq:persupp}) implies that
\begin{equation} \label{eq:specmin} E_0 := \min \Sigma = \inf \{
  \min\,\sigma(H_{\omega}): \omega \in {\mathcal C}_{per}\}.
\end{equation}

It is a non-trivial question to decide if there is a periodic
minimizer, i.e.\ if the infimum in (\ref{eq:specmin}) is a minimum. In
fact, we do not believe that this is true in general. Our choice of
the following assumptions on $q$ and $\mu$ is mostly motivated by the
fact that they allow to find a periodic minimizer for
(\ref{eq:specmin}).

\vspace{.3cm}

{\bf (A1)} The single-site potential $q:\R^d \to \R$ is bounded,
measurable and reflection-symmetric in each variable. Moreover,
supp$\,q \subset [-r,r]^d$ for some $r<1/2$.

\vspace{.3cm}

{\bf (A2)} Let $d_{max} := \frac{1}{2}-r$ and ${\mathcal C} := \{(\pm
d_{max}, \ldots, \pm d_{max})\}$ denote the $2^d$ corners of the
closure $\overline{G}$ of $G:=(-d_{max}, d_{max})^d$. Then
\[ {\mathcal C} \subset \: \mbox{supp}\,\mu \subset \overline{G}.\]

\vspace{.3cm}

\begin{figure}[ht]
  \centering
  \includegraphics[width=5cm]{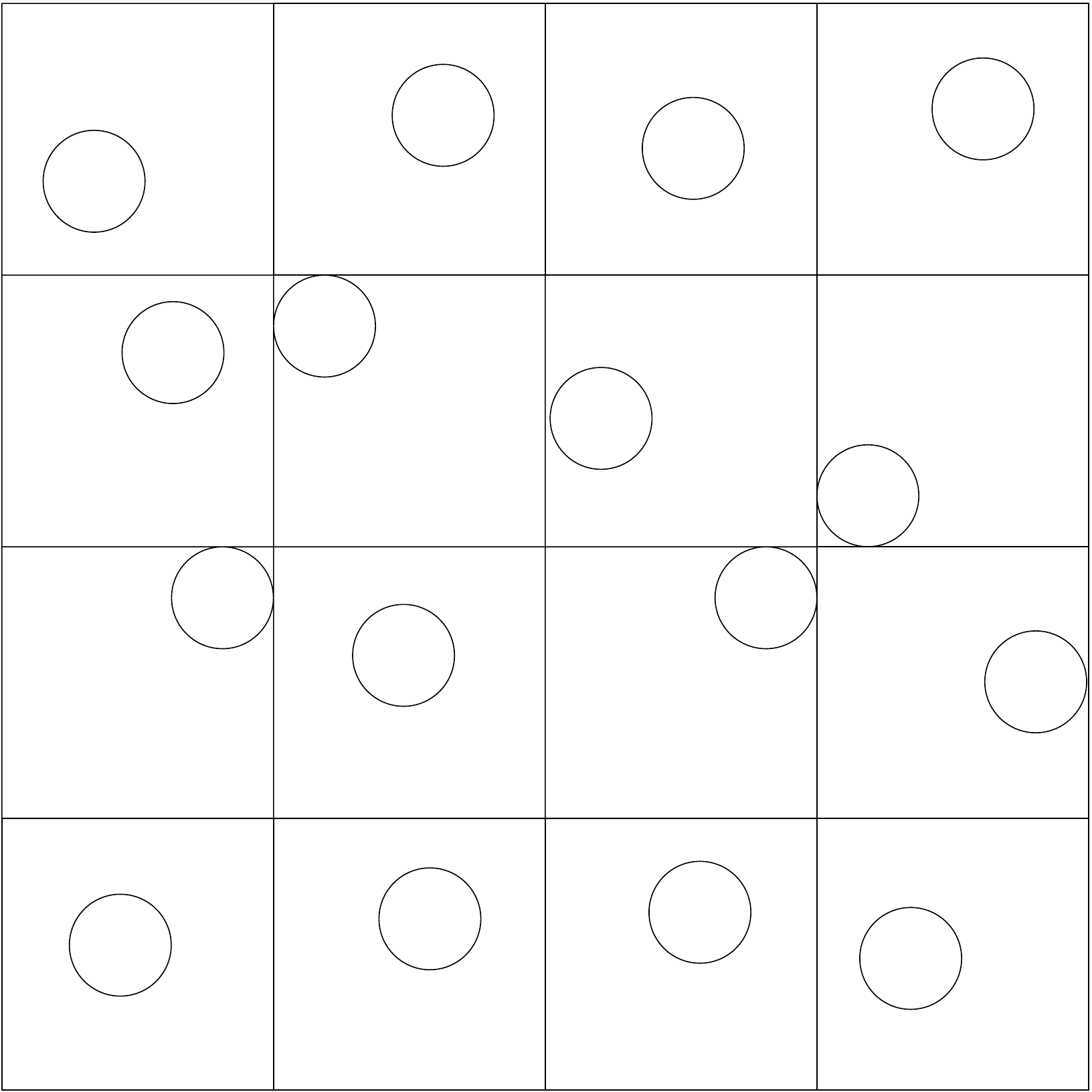}
  \caption{The support of $V_{\omega}$ for a typical $\omega$.}
  \label{fig1}
\end{figure}

The two support assumptions on $q$ and $\mu$ have a simple geometric
interpretation for the RDM (\ref{eq:RDM}): The support of each
single-site term $q(\cdot-n-\omega_n)$ stays in the unit cell centered
at $n$, while it is allowed to ``touch'' the boundary of the cell. In
fact, with positive probability the single-site potentials may move
arbitrarily close to each corner of their cell. For a typical
configuration of the $\omega_n$ see Figure~\ref{fig1} (where the
support of $q$ is drawn radially symmetric for aesthetic reasons).

We can now identify a periodic minimizer for (\ref{eq:specmin}),
stating a result from \cite{BLS1}:

\begin{theorem} \label{thm1} Assume {\bf (A1)} and {\bf (A2)} and let
  $\omega^* = (\omega_n^*)_{n\in \Z^d}$ be given by
  \begin{equation} \label{eq:minimizer} \omega_n^* := ((-1)^{n_1}
    d_{max}, \ldots, (-1)^{n_d} d_{max}), \quad n = (n_1, \ldots, n_d)
    \in \Z^d.
  \end{equation}
  Then $E_0 = \min \sigma(H_{\omega^*})$.
\end{theorem}

\begin{figure}[h]
  \centering
  \includegraphics[width=5cm]{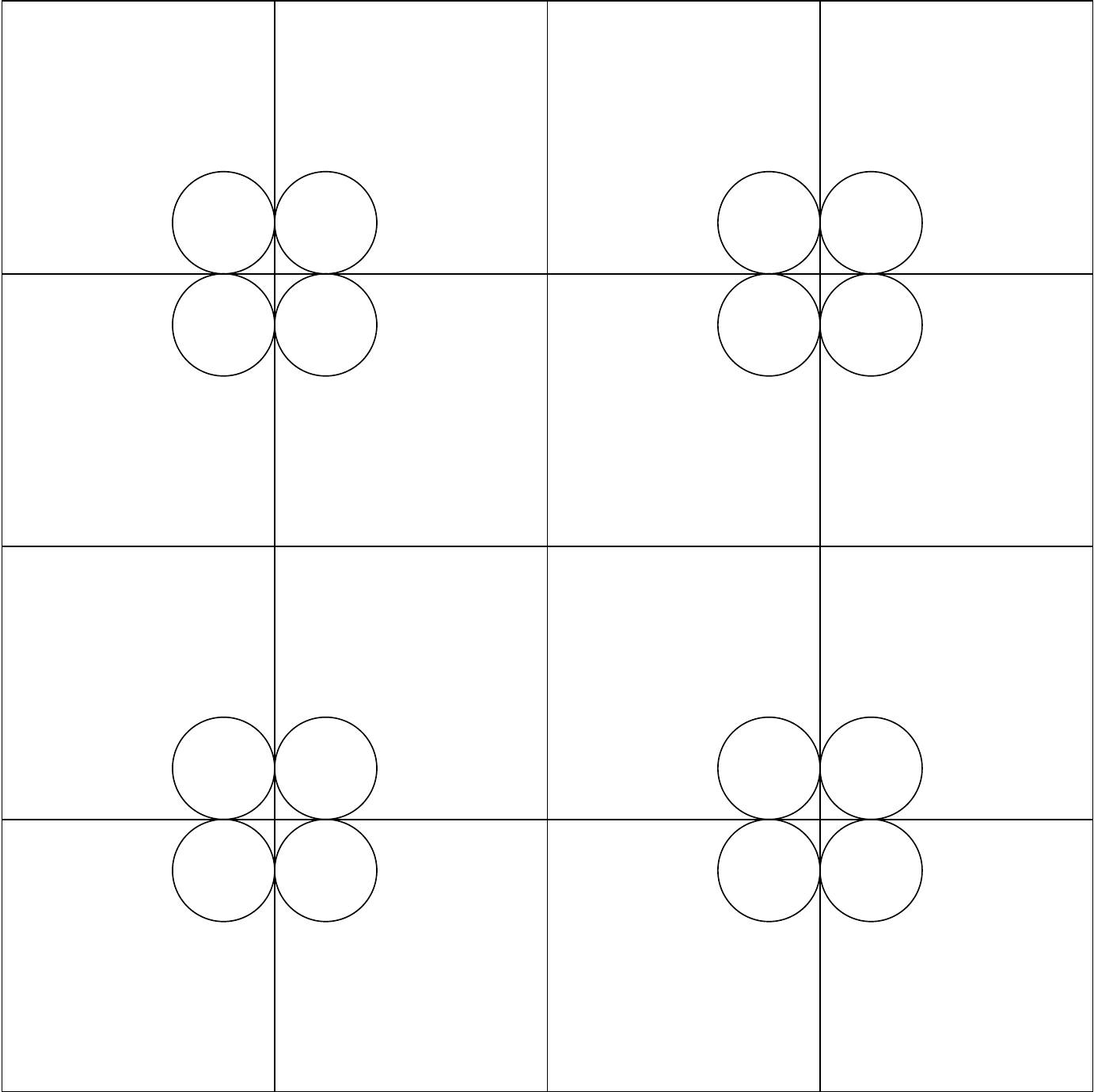}
  \caption{Support of $V_{\omega^*}$ for $d=2$.}
  \label{fig2}
\end{figure}

The potential $V_{\omega^*}(x)= \sum_n q(x-n-\omega_n^*)$ is
2-periodic in each direction and locally consists of densest clusters
of $2^d$ single-site terms placed into adjacent corners of their
cells, see Figure~\ref{fig2}.

One may understand this result heuristically by the following strategy
to construct test-functions which minimize the quadratic form of
$H_{\omega}$, at least if the single-site potential $q$ is negative:
The clusters in $V_{\omega^*}$ form wide wells. In these wells one can
place localized test functions with relatively small derivative, due
to the width of the wells, i.e.\ small cost in kinetic energy. This
gives lower total energy than the narrower wells given by individual,
spatially separated single-site potentials. This is not how
Theorem~\ref{thm1} is proven. We should also point out that
Theorem~\ref{thm1} does not impose any sign-restrictions on $q$, and
thus can not be fully explained by the above heuristics. But the
heuristics make clear that the spectral minimum of the RDM is
determined by a non-trivial interplay between kinetic and potential
energy.

Instead we will give a proof of Theorem~\ref{thm1} at the end of the
next section, based on the answer to a minimization problem for a
single-site Neumann operator associated with the RDM.

\section{The Neumann Problem} \label{sec:Neumann}

Theorem~\ref{thm1} provides the answer to an optimization problem
involving infinitely many displacement parameters $\omega_n$, $n\in
\Z^d$. However, due to the symmetry assumptions on $q$, it turns out
that the proof can be reduced to a related problem involving the
optimal placement of just one single-site term.

For this purpose, let $\Lambda_1 := (-\frac{1}{2}, \frac{1}{2})^d$ be
the unit cube centered at the origin and $-\Delta^N$ the
Neumann-Laplacian on $L^2(\Lambda_1)$, i.e.\ the unique self-adjoint
operator whose quadratic form is $\int_{\Lambda_1} |\nabla
f(x)|^2\,dx$ for $f \in H^1(\Lambda_1)$, the first order Sobolev
space.

For $q$ as in {\bf (A1)} and $a\in \overline{G}$ let
\[ H_{\Lambda_1}^N(a) := -\Delta^N + q(x-a)\] and $E_0(a) := \min
\sigma(H_{\Lambda_1}^N(a))$ the non-degenerate lowest eigenvalue of
$H^N_{\Lambda_1}(a)$. For a general discussion of properties of
operators of this type see Section 2 of \cite{BLS1}.

We ask for the optimal placement of $a\in \overline{G}$ to minimize
$E_0(a)$ and arrive at the following result.

\begin{theorem} \label{thm2}
Under assumption {\bf(A1)} one of the
  following two alternatives holds:

  (i) $E_0(a)$ is strictly maximized at $a=0$ and strictly minimized
  at the corners ${\mathcal C}$ of $\overline{G}$.

  (ii) $E_0(a)$ is identically zero.

\end{theorem}

A proof of Theorem~\ref{thm2} under the given assumptions can be found
in \cite{BLS1}. We will not discuss details of this proof here, as we
will later need a strengthened version of Theorem~\ref{thm2} for which
we will also use somewhat stronger assumptions, see Theorem~\ref{thm7}
and assumption {\bf (A1)'} in Section~\ref{sec:missinglink} below.

The proof of Theorem~\ref{thm2} in \cite{BLS1} shows that in each of
the two alternatives more can be said:

In case of alternative (i), the function $E_0(a_1, \ldots, a_j,
\ldots, a_d)$ is symmetric and strictly unimodal in each
variable. Thus, with all other variables fixed, for each $j$ it is a
strictly increasing function of $a_j$ in $[-d_{max},0]$ and strictly
decreasing in $[0,d_{max}]$.

On the other hand, if alternative (ii) holds, then the strictly
positive ground state eigenfunction $u_0(x,a)$ corresponding to
$E_0(a)$ is constant near the boundary of $\Lambda_1$ (and thus, by
analyticity, constant in the entire connected component of $\Lambda_1
\setminus \mbox{supp}\,q(\cdot-a)$ containing the boundary of
$\Lambda_1$). This reveals a mechanism which can be used to construct
non-trivial examples (with non-vanishing $q$) where alternative (ii)
happens:

Let $\phi(x)$ be a positive sufficiently regular function which is
constant near the boundary of $\Lambda_1$ and then define the
potential by setting
\begin{equation} \label{eq:iiexample} q(x-a) = \frac{\Delta
    \phi(x-a)}{\phi(x-a)}
\end{equation}
as long as supp$\,q(\cdot - a) \subset \Lambda_1$. Then $E_0(a)$ vanishes
identically for $a\in G$ and $\phi(x-a)$ is the corresponding
eigenfunction. As follows from the proof of Theorem~\ref{thm2}, this
is the only mechanism which leads to alternative (ii).

Alternative (i) certainly happens for all non-vanishing sign-definite
potentials $q$, as it follows by perturbation theory that in this case
the zero ground state energy $0$ of the Neumann Laplacian is pushed
either up or down. But alternative (i) is generic also for
sign-indefinite potentials, as alternative (ii) will be broken by
typical small perturbations of the potential.

Among previously known results, the ones most closely related to
Theorem~\ref{thm1} can be found in \cite{Harrelletal} which considers
similar questions for the case of the Dirichlet Laplacian $-\Delta^D$
instead of $-\Delta^N$. Also using symmetry assumptions on $q$, it is
found there that the optimal placement of the potential in $-\Delta^D
+ q(x-a)$ depends strongly on the sign of $q$. For cubic domains, a
special case of the domains considered in \cite{Harrelletal}, it is
found that for positive potential the lowest eigenvalue is minimized
if the potential is placed in a corner of the cube, while negative
potentials should be placed into the center of the cube. This
distinction does not happen in the Neumann case, where it is generally
true that ``bubbles tend to the corners''.

While not used in our proof of Theorem~\ref{thm2} or in the proofs in
\cite{Harrelletal}, one can understand this distinction by
perturbative arguments. For this, consider $-\Delta + \lambda q(x-a)$
on $L^2(\Lambda_1)$ for small coupling, with either Dirichlet or
Neumann boundary condition. If $E_0^D(a, \lambda)$ denotes the
smallest eigenvalue in the Dirichlet case, then by first order
perturbation theory,
\begin{equation} \label{eq:firstorder}
  \partial_{\lambda} E_0^D(a,0) = \int q(x-a) |\varphi(x)|^2\,dx,
\end{equation}
where $\varphi(x)$ is the normalized ground state of the Dirichlet
Laplacian on $\Lambda_1$, i.e.\ $\varphi(x) = C \prod_{j=1}^d \cos
(\pi x_j)$. In the small coupling regime minimizing
(\ref{eq:firstorder}) over $a$ indicates the optimal placement of the
potential. If $q$ is positive, then the bubble should be placed into a
corner of $\Lambda_1$, where $|\varphi|^2$ has the smallest mass. On
the other hand, for negative $q$ the bubble should be placed into the
center where the mass of $|\varphi|^2$ is largest.

For the Neumann case the heuristics given by first order perturbation
theory is inconclusive. The ground state of the Neumann Laplacian is
constant, and thus $\partial_{\lambda} E_0^N(a,0)$ is independent of
$a$.

However, one gets correct heuristics by going to second order
perturbation theory. We have (for a derivation see Section~2.3 of
\cite{BLS1})
\begin{equation} \label{eq:secondorder}
  \partial_{\lambda}^2 E_0^N(a,0) = - 2 \sum_{k>0} \frac{(u_0, q(\cdot-a) u_k)^2}{E_k - E_0}.
\end{equation}
Here $0 = E_0 < E_1 \le E_2 \le \ldots$ are the eigenvalues of the
Neumann Laplacian and $u_k$ the corresponding eigenfunctions. In $d=2$
(for simplicity) we have that the first excited state is twice
degenerate, $E_1 = E_2 = \pi^2$. Considering only these two terms in
(\ref{eq:secondorder}) (the third term would still give the same
result) we get that $\partial_{\lambda}^2 E_0^N(a,0)$ is approximately
given by
\[ -\frac{4}{\pi^2} \left[ \left( \int q(x-a_1, y-a_2) \sin(\pi x)
    \,dx\,dy \right)^2 + \left( \int q(x-a_1,y-a_2) \sin(\pi y)
    \,dx\,dy \right)^2 \right], \] which is non-positive. If $q$ is
reflection symmetric, then both integrals are zero for $a=0$,
indicating the position with highest ground state energy in the small
coupling regime. If we also assume that $q$ is of fixed sign, then
both integrals become maximal in absolute value if $a$ is located near
one of the corners of the cube. These are the positions where the
ground state energy of $-\Delta^N +\lambda q(x-a)$, $\lambda \approx
0$, is minimal. As opposed to the Dirichlet case, the answer suggested
by second order perturbation theory is the same for positive and
negative $q$.

Let us finally start to get beyond heuristics and show rigorously that
Theorem~\ref{thm2} implies Theorem~\ref{thm1}:

\begin{proof} (of Theorem \ref{thm1}, \cite{BLS1}) For any given
  configuration $\omega$, the restriction of $H_{\omega}$ to the unit
  cube centered at $n\in \Z^d$ with Neumann boundary conditions is
  unitarily equivalent (via translation by $n$) to
  $H_{\Lambda_1}^N(\omega_n)$, defined as in Theorem~\ref{thm2}. Thus,
  by Neumann bracketing and Theorem~\ref{thm2},
  \begin{eqnarray*}
    \min \sigma( H_{\omega} ) &\geq& \min \sigma \left(
      \bigoplus_{n \in \Z^d} H^N_{\Lambda_1}(\omega_n) \right) \\
    &\geq& \inf \left\{E_0(a): \,a\in [-d_{max},d_{max}]^d\right\} \\
    &=& E_0(a^*),
  \end{eqnarray*}
  where $a^* = (d_{max},\ldots, d_{max})$ is one of the corners
  ${\mathcal C}$ of $\overline{G}$.  This holds for arbitrary
  configurations $\omega$ and thus, by \eqref{eq:asspectrum}, $E_0 =
  \min \Sigma \ge E_0(a^*)$.

  Now consider $\omega^* = (\omega_n^*)_{n\in \Z^d}$ as given by
  (\ref{eq:minimizer}). The corresponding potential $V_{\omega^*}(x) =
  \sum_{n\in \Z^d} q(x-n-\omega_n^*)$ is $2$-periodic in $x_j$ for
  each $j$. By Floquet-Bloch theory \cite{Barry4} the bottom of the
  spectrum of $H_{\omega^*} = -\Delta + V_{\omega^*}$ is given by the
  smallest eigenvalue $E_0^{per}$ of its restriction to $\Lambda_0^2
  := (-\frac{1}{2}, \frac{3}{2})^d$ with periodic boundary conditions,
  see Figure~\ref{fig:periodtwomin}.

\begin{figure}[h]
  \centering
  \includegraphics[width=5cm]{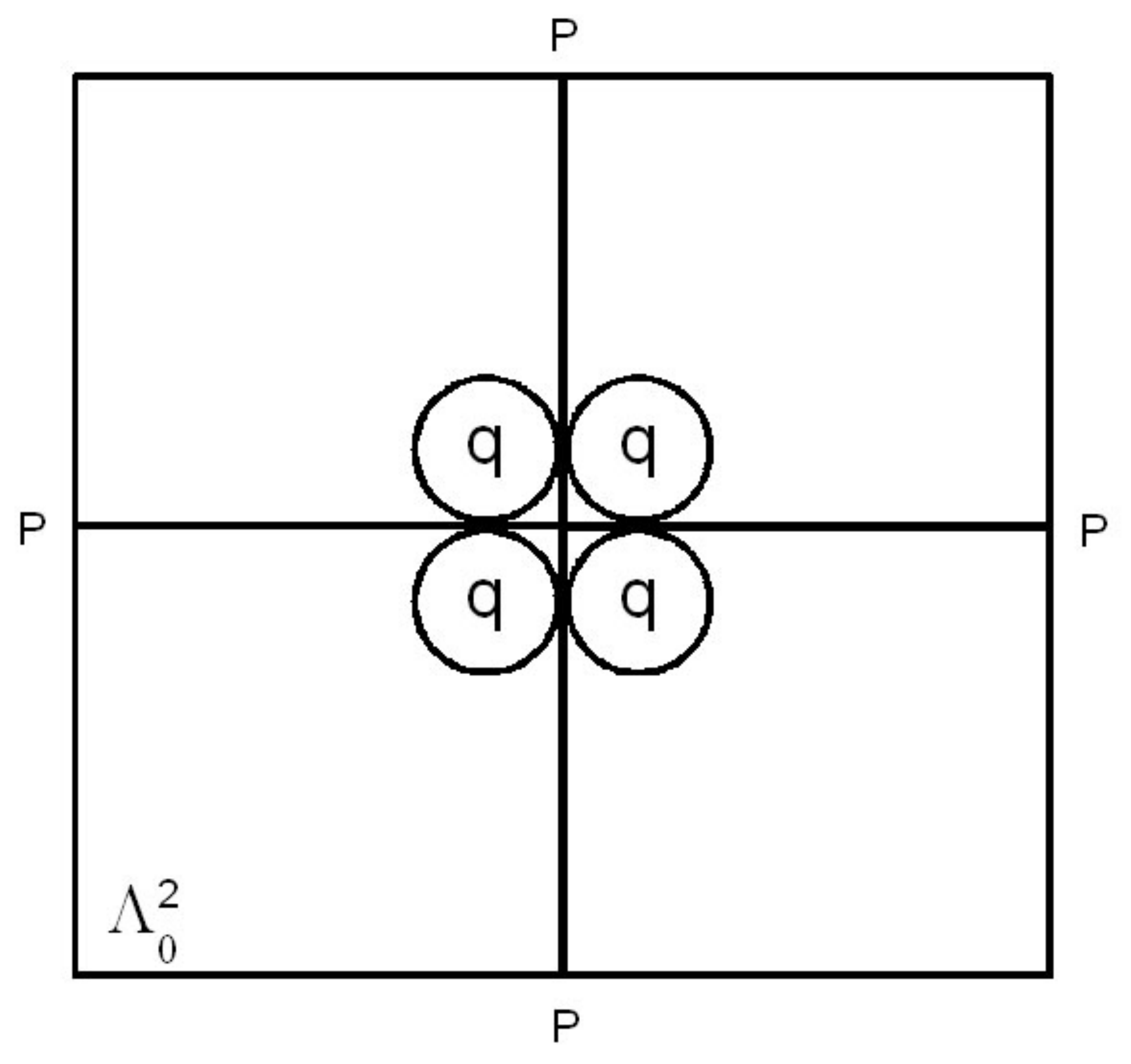}
  \caption{The period cell of $V_{\omega^*}$ in $d=2$.}
  \label{fig:periodtwomin}
\end{figure}

On $\Lambda_0^2$ the potential $V_{\omega^*}$ is symmetric with
respect to all hyperplanes $x_i=1/2$, $i=1,\ldots,d$. Thus $E_0^{per}$
coincides with the smallest eigenvalue of the Neumann problem on
$\Lambda_0^2$. Again by symmetry of the potential, the latter
coincides with the smallest eigenvalue of the Neumann problem on
$\Lambda_1$. As $\omega_0^* = a^*$, this eigenvalue is
$E_0(a^*)$. Together with (\ref{eq:specmin}) we have shown that
\[ E_0 \le \min \sigma(H_{\omega^*}) = E_0(a^*). \] Combined with the
previous observation that $E_0(a^*) \le E_0$ this shows $E_0 = \min
\sigma (H_{\omega^*})$.
\end{proof}

\section{Uniqueness of the Periodic Minimizer} \label{sec:uniqueness}

After resolving the preliminary problem of characterizing the spectral
minimum of the RDM, we could now turn to the other essential
ingredients into a localization proof, a Lifshitz tail bound on the
IDS and a Wegner estimate. However, a first look at this quickly
demonstrates that we also need to address the question of uniqueness
in Theorem~\ref{thm1}. Other than translates of $\omega^*$, are there
more periodic configurations which have the same spectral minimum?

To motivate this, let us include a first discussion of Lifshitz
tails. For this one considers restrictions $H_{\omega,L}$ of
$H_{\omega}$ to $L^2(\Lambda_L)$, where $\Lambda_L$ is a cube of
side-length $L$ centered at the origin. As boundary condition one can
generally choose what is most convenient in a given model, for us this
will be Neumann conditions. By a Lifshitz tail bound we mean a result
which says that the probability of $H_{\omega,L}$ to have an
eigenvalue close to $E_0$, the minimum of the infinite volume
spectrum, is exponentially small in $L$. The meaning of ``close'' will
be made more precise later.

If $\omega$ coincides with $\omega^*$ on $\Lambda_L$ or is very close
to it, this will give a low lying eigenvalue of $H_{\omega,L}$. Our
chances of getting a useful Lifshitz tail bound would worsen if there
are many other periodic configurations with the same spectral minimum
as $\omega^*$, as this would increase the probability that random
configurations are close to one of the minimizing configurations on
$\Lambda_L$ and thus have low lying eigenvalues.

From this it is immediately clear that for all further considerations
we will have to assume that alternative (i) of Theorem~\ref{thm2}
holds, as under alternative (ii) it follows that $H_{\omega,L}$ has
spectral minimum $E_0$ for {\it every} configuration $\omega$. The
following result is taken from \cite{BLS2}.

\begin{theorem} \label{thm3} Assume {\bf (A1)}, {\bf (A2)},
  alternative (i) of Theorem~\ref{thm2}, $d\ge 2$ and $r<1/4$. Then
  $\omega^*$ as given by (\ref{eq:minimizer}) is, up to translations,
  the unique periodic configuration with $\min \sigma(H_{\omega^*}) =
  E_0$.
\end{theorem}

Two additional assumptions were made here which deserve comment: For
the ``radius'' $r$ of the single-site potential $q$ we require $r<1/4$
rather then just $r<1/2$ assumed earlier. This is a technical
assumption, which we need to apply an analyticity argument in the
proof, see below. Our guess is that this assumption is not necessary
for Theorem~\ref{thm3} to hold.

However, Theorem~\ref{thm3} indeed only holds in the multi-dimensional
case $d\ge 2$ In the case $d=1$ there are many periodic minimizers, as
also proven in \cite{BLS2}:

\begin{theorem} \label{thm4} Assume {\bf (A1)}, {\bf (A2)},
  alternative (i) of Theorem~\ref{thm2} and $d=1$. Then an
  $L$-periodic configuration $\omega = (\omega_n)_{n\in \Z}$,
  $\omega_{n+L}= \omega_n$ for all $n\in \Z$, satisfies $\min
  \sigma(H_{\omega}) = E_0$ if and only if

  (i) all $\omega_n$ are maximally displaced, i.e.\ $\omega_n = \pm
  d_{min}$ for all $n$,

  (ii) $L$ is even, and

  (iii) in each period $L$ equally many $\omega_n$ are displaced to
  the left and to the right.
\end{theorem}

\begin{figure}[h]
  \centering
  \includegraphics[width=4in]{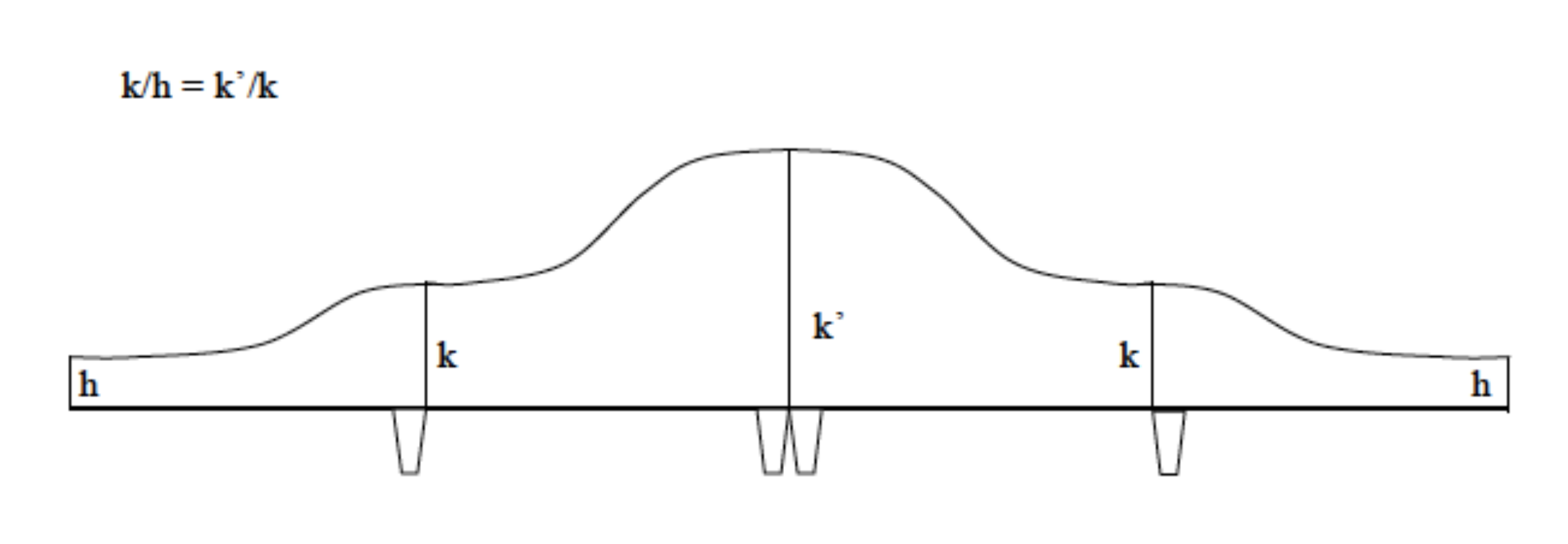}
  \caption{A 4-periodic minimizer in $d=1$.}
  \label{fig:4}
\end{figure}

It is easy to see that a periodic configuration $\omega$ with these
properties is a minimizer. Let $\varphi_0$ be the positive ground
state of $-d^2/dx^2 + q(x-d_{max})$ on $(-1/2, 1/2)$ with Neumann
boundary conditions. It can be shown that alternative (i) implies that
\[ h:= \varphi_0(-1/2) \not= \varphi_0(1/2) =: k.\] For a
configuration satisfying (i), (ii) and (iii) of Theorem~\ref{thm4} the
Neumann ground state over the period $L$ is found by pasting together
scaled copies of $\varphi_0$, compare Figure~\ref{fig:4} for an
example with $L=4$. The number of steps up is equal to the number of
steps down, which allows for periodic extension, showing that $\min
\sigma(H_{\omega}) = E_0$.

Lemma~\ref{lem1} below, which holds in arbitrary dimension,
establishes the necessity of (i). The above construction of the
Neumann ground state for the period $L$ now shows that (ii) and (iii)
must hold for the Neumann ground state to coincide with the periodic
ground state. For a more detailed proof of Theorem~\ref{thm4} see
\cite{BLS2}.

Theorem~\ref{thm4} has surprising implications for the integrated
density of states $N(E)$ of the one-dimensional RDM. The most extreme
situation occurs if $\mu$ is the Bernoulli measure with equal weights
at the endpoints of $[-d_{max}, d_{max}]$, i.e.\
\begin{equation} \label{eq:Bernoulli} \mu = \frac{1}{2} \delta_{d_{max}}
  + \frac{1}{2} \delta_{-d_{max}}.
\end{equation}

\begin{theorem} \label{thm5} Let $H_{\omega}$ be the one-dimensional
  RDM with distribution $\mu$ given by (\ref{eq:Bernoulli}). Then
  there exists $C>0$ such that
  \begin{equation} \label{eq:IDS} N(E) \ge \frac{C}{(\log(E-E_0))^2}
  \end{equation}
  for $E$ sufficiently close to $E_0$.
\end{theorem}

While similar phenomena have been found for Schr{\"o}dinger operators with
almost-periodic potentials, this is the first known example of a
random Schr{\"o}dinger operator with non-H{\"o}lder-continuous IDS. The
density $n(E)=N'(E)$ of eigenvalues near the bottom of the spectrum is
even higher than for the one-dimensional Laplacian where the IDS has a
square-root type singularity $N(E) = CE^{1/2}$ at $E_0=0$. Thus the
randomness has the effect of pulling more eigenvalues towards the
bottom of the spectrum, rather than pushing them away from the bottom
as in the more common fluctuation boundary regime described by
Lifshitz tails. The reason behind (\ref{eq:IDS}) is Theorem~\ref{thm4}
combined with the law of large numbers. For the symmetric Bernoulli
distribution (\ref{eq:Bernoulli}) it has very high probability that in
a large even period $L$ almost equally many $\omega_n$ take values
$d_{max}$ and $-d_{max}$, leading to a ground state energy very close
to $E_0$. For a detailed proof of Theorem~\ref{thm5} see \cite{BLS2}.

We now turn back to the original goal of this section, the proof of
Theorem~\ref{thm3} on the uniqueness of the periodic minimizer in
$d\ge 2$. For this we consider a configuration $\omega \in {\mathcal
  C}_{per}$ (as defined in (\ref{eq:perconfig})) and let $\Lambda$ be
the corresponding rectangular period cell. We let
$H_{\omega,\Lambda}^P$ and $H_{\omega,\Lambda}^N$ be the restriction
of $H_{\omega}$ to $L^2(\Lambda)$ with periodic and Neumann boundary
conditions, respectively, and $E_0(H_{\omega,\Lambda}^P)$ and
$E_0(H_{\omega,\Lambda}^N)$ their lowest eigenvalues. In follows from
general facts that
\begin{equation} \label{eq:floquetfacts} \min \sigma(H_{\omega}) =
  E_0(H_{\omega,\Lambda}^P) \ge E_0(H_{\omega,\Lambda}^N).
\end{equation}
We assume that $\min \sigma(H_{\omega}) = E_0$ and have to show that,
up to a translation, $\omega$ coincides with $\omega^*$. This is done
in two steps.

The first step establishes that all $\omega_n$ sit in corners and that
the ground state of $H_{\omega,\Lambda}^N$ satisfies Neumann
conditions not only on $\Lambda$, but on every unit cell contained in
$\Lambda$:

\begin{lemma} \label{lem1} Let $\omega$ be a periodic configuration
  with $\min \sigma(H_{\omega}) =E_0$. Then $\omega_n \in {\mathcal
    C}$ for all $n\in \Z^d$.  Moreover, in this case
  $E_0(H_{\omega,\Lambda}^P) = E_0(H_{\omega,\Lambda}^N)$ and the
  ground state eigenfunction $\psi_{\omega}$ of $H_{\omega,\Lambda}^N$
  satisfies Neumann boundary conditions on the boundary of each unit
  cube $\Lambda_n$ centered at $n\in \Lambda \cap \Z^d$.
\end{lemma}

The core of the proof of Lemma~\ref{lem1} is the following
calculation, based on Neumann bracketing and the characterization of
ground state energies as minimizers of the quadratic form:

\begin{eqnarray} \label{eq:Neumannbrack}
  E_0(H_{\omega,\Lambda}^N) & = & \frac{\int_{\Lambda} |\nabla \psi_{\omega}|^2 + \int_{\Lambda} \sum_{n\in \Lambda \cap \Z^d} q(x-n-\omega_n) |\psi_{\omega}|^2}{\int_{\Lambda} |\psi_{\omega}|^2} \nonumber \\
  & = & \sum_{n\in \Lambda \cap \Z^d} \frac{\int_{\Lambda_n} |\nabla \psi_{\omega}|^2 + \int_{\Lambda_n} q(x-n-\omega_n) |\psi_{\omega}|^2}{ \int_{\Lambda_n} |\psi_{\omega}|^2} \cdot \frac{\int_{\Lambda_n} |\psi_{\omega}|^2}{\int_{\Lambda} |\psi_{\omega}|^2} \nonumber \\
  & \ge & \sum_{n\in \Lambda \cap \Z^d} E_0(\omega_n)
  \frac{\int_{\Lambda_n} |\psi_{\omega}|^2}{\int_{\Lambda}
    |\psi_{\omega}|^2} \ge \sum_{n\in \Lambda \cap \Z^d} E_0
  \frac{\int_{\Lambda_n} |\psi_{\omega}|^2}{\int_{\Lambda}
    |\psi_{\omega}|^2} = E_0,
\end{eqnarray}

By (\ref{eq:floquetfacts}) and the assumption we conclude that
$E_0(H_{\omega,\Lambda}^P) = E_0(H_{\omega,\Lambda}^N)$ and that all
inequalities in (\ref{eq:Neumannbrack}) are equalities. We also see
that $\omega_n \in {\mathcal C}$ for all $n\in \Z^d$, because
otherwise, given alternative (i), the last inequality in
(\ref{eq:Neumannbrack}) would be strict. Finally we see from equality
in the second to last inequality in (\ref{eq:Neumannbrack}) that
$\psi_{\omega}|_{\Lambda_n}$ is the ground state for the Neumann
problem on $\Lambda_n$, and thus satisfies Neumann conditions on
$\Lambda_n$.

The second step of the proof of Theorem~\ref{thm3} is to show
symmetric matching of the bubbles, i.e.\ that in each pair of
neighboring unit cells within $\Lambda$ the single-site potentials are
placed symmetrically with respect to the common boundary of the
cells. For this we use the following general fact from \cite{BLS2}, to
where we refer for the proof:

\begin{lemma} \label{lem2} Consider a connected open region $D$ in
  $\R^d$, $d\ge 2$ and a hyperplane $P$ that divides this region into
  two nonempty subregions. Denote by $\sigma$ the reflection about $P$
  and assume that $D\cap \sigma(D)$ is connected. Let $E\in \R$ and,
  in $D$, let $u$ be a solution of the equation
  \begin{equation} \label{schroequ} -\Delta u = E u
  \end{equation}
  which satisfies the condition $ \frac{\partial u}{\partial n} = 0$
  on $P\cap D $.  Then $u$ can be extended to a symmetric function $w$
  on $D \cup \sigma (D)$ which satisfies the equation $-\Delta u = E
  u$ in this region.
\end{lemma}

\begin{figure}[h]
  \begin{center}
    \includegraphics[width=2.5in]{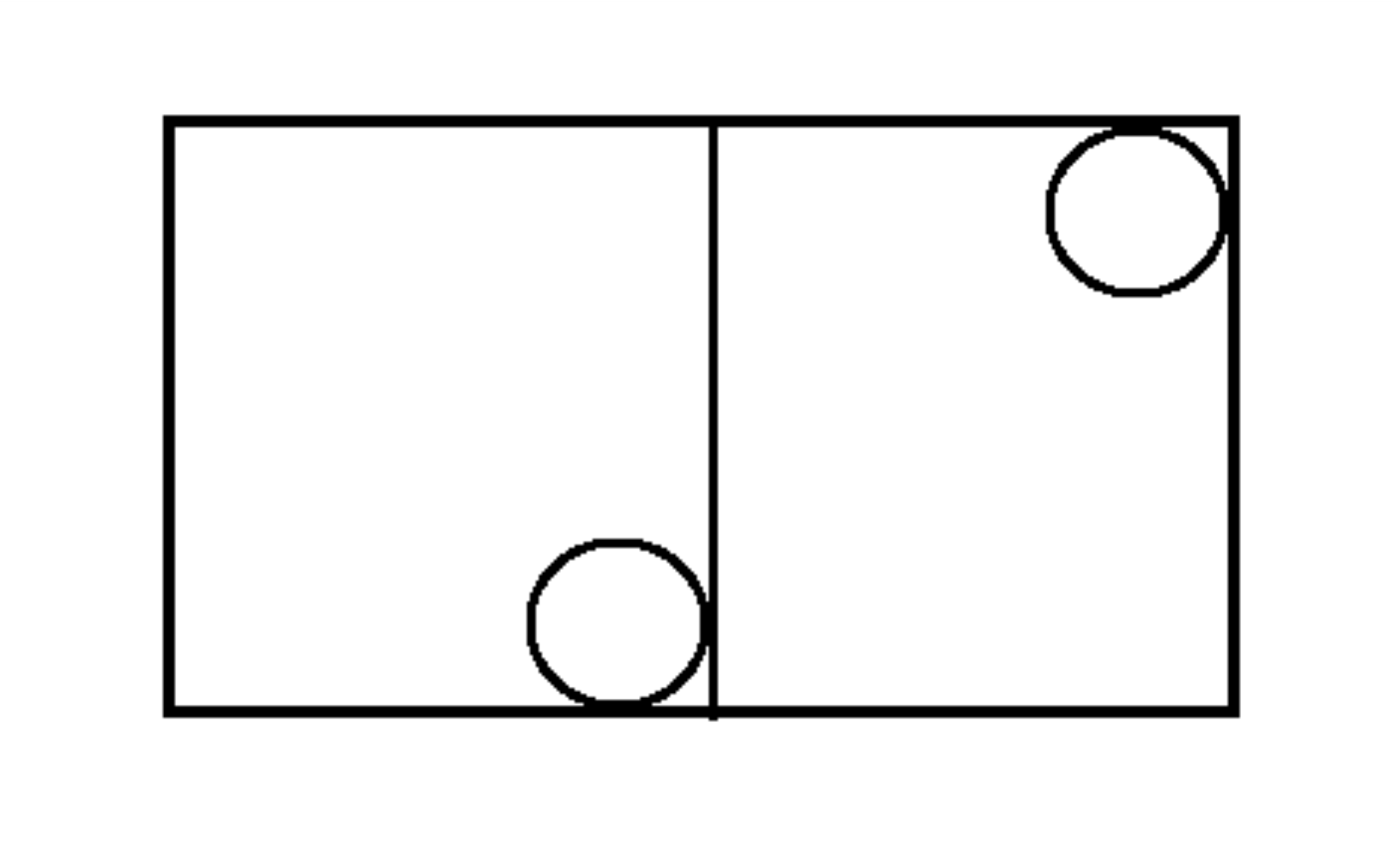}
  \end{center}
  \caption{A non-matching pair.}
  \label{fig4}
\end{figure}

To finish the proof of Theorem~\ref{thm3} let us assume that $\omega$
is a periodic minimizing configuration in which, by Lemma~\ref{lem1},
all bubbles sit in corners, but that there is at least one
non-matching neighboring pair of bubbles. Let us focus on $d=2$ and
the situation in Figure~\ref{fig4} (the general argument in
\cite{BLS2} uses the same idea).
Circumscribe squares around the
supports of the two bubbles and remove these two squares from the
union $R$ of the two cells. Choose the resulting region as $D$ in
Lemma~\ref{lem2}, and $\sigma$ as reflection at the center line. Here
(and only here) we need the strengthened assumption $r<1/4$ in
Theorem~\ref{thm3} to make sure that $D \cap \sigma(D)$ is connected.

Let $u$ be the restriction of the ground state of
$H_{\omega,\Lambda}^N$ to $D$. Then $-\Delta u =E_0 u$ on $D$ and, by
Lemma~\ref{lem1}, $u$ satisfies Neumann conditions on the centerline
$P\cap D$. By Lemma~\ref{lem2} $u$ can be extended to a symmetric
function $w$ on $D \cup \sigma(D)$ satisfying $-\Delta w = E_0 w$. But
$D \cup \sigma(D) = R$ and thus $w$ is the ground state of the Neumann
Laplacian on $R$, implying $E_0=0$ and that $w$ is constant, a
contradiction to alternative (i). This completes the proof of
Theorem~\ref{thm3}.

To end this section let us remark that the fundamental difference in
the one-dimensional and multi-dimensional case lies in the possibility
to smoothly match Neumann ground states on different unit cells. In
$d=1$ this can always be done by re-scaling as explained by
Figure~\ref{fig4}. In higher dimension the boundary of cells has much
more structure (is not just a point), which ultimately results in
matching of ground states only being possible in the trivial
reflection-symmetric case.

\section{Special Lifshitz tails} \label{sec:SpecLifshitz}

After clarifying uniqueness questions in the previous section we
finally have enough background information to enter into a discussion
of Lifshitz tail properties of the IDS for the random displacement
model. In order to use our earlier results we will have to assume from
here on that alternative (i) of Theorem~\ref{thm2} holds and that
$d\ge 2$. We strengthen {\bf (A2)} to require $r<1/4$ and will in this
section also assume that $q$ is continuous to make use of results in
\cite{KN2}.

Let $H_{\omega,L}^N$ be the restriction of $H_{\omega}$ to $\Lambda_L
= (-L-1/2,L+1/2)^d$ with Neumann boundary conditions. The crucial fact
required in localization proofs and also in the proof of Lifshitz tail
asymptotics of the IDS is that the probability of $H_{\omega,L}^N$
having an eigenvalue close to $E_0$ is very small. The first proof of
this was given in \cite{KN2}, where it follows as a special case of a
more general result. The methods derived in \cite{KN2} require to
assume in {\bf (A2)} that supp$\,\mu$ is {\it finite}. The methods
developed later in \cite{KLNS}, which are described in
Section~\ref{sec:GenLifshitz} below, have allowed to remove this
additional assumption on $\mu$. However, Theorem~\ref{thm6} is crucial
as it will serve as the anchor for a bootstrap argument in
Section~\ref{sec:GenLifshitz}. For this it will be sufficient to start
with the case supp$\,\mu = {\mathcal C}$, i.e.\ a displacement model
where all bubbles sit in corners.

\begin{theorem} \label{thm6} Let the assumptions listed at the
  beginning of this section be satisfied and also assume that
  supp$\,\mu= {\mathcal C}$. Then there exist $C>0$ and $\mu>1$ such
  that for all $L\in \N$,
  \begin{equation} \label{eq:Lifshitzbound} \PP\left( \min
      \sigma(H_{\omega,L}^N) < E_0 + \frac{C}{L^2} \right) <
    (2L+1)^{d-1} \mu^{-2L}.
  \end{equation}
\end{theorem}

In the remainder of this section we will discuss the argument from
\cite{KN2} which proves Theorem~\ref{thm6}, taking some advantage in
presentation from only looking at the specific situation which is of
interest to us here. However, we will refer to \cite{KN2} for many of
the core analytical parts of the proof.

\begin{figure}[h]
  \begin{center}
    \includegraphics[width=3in]{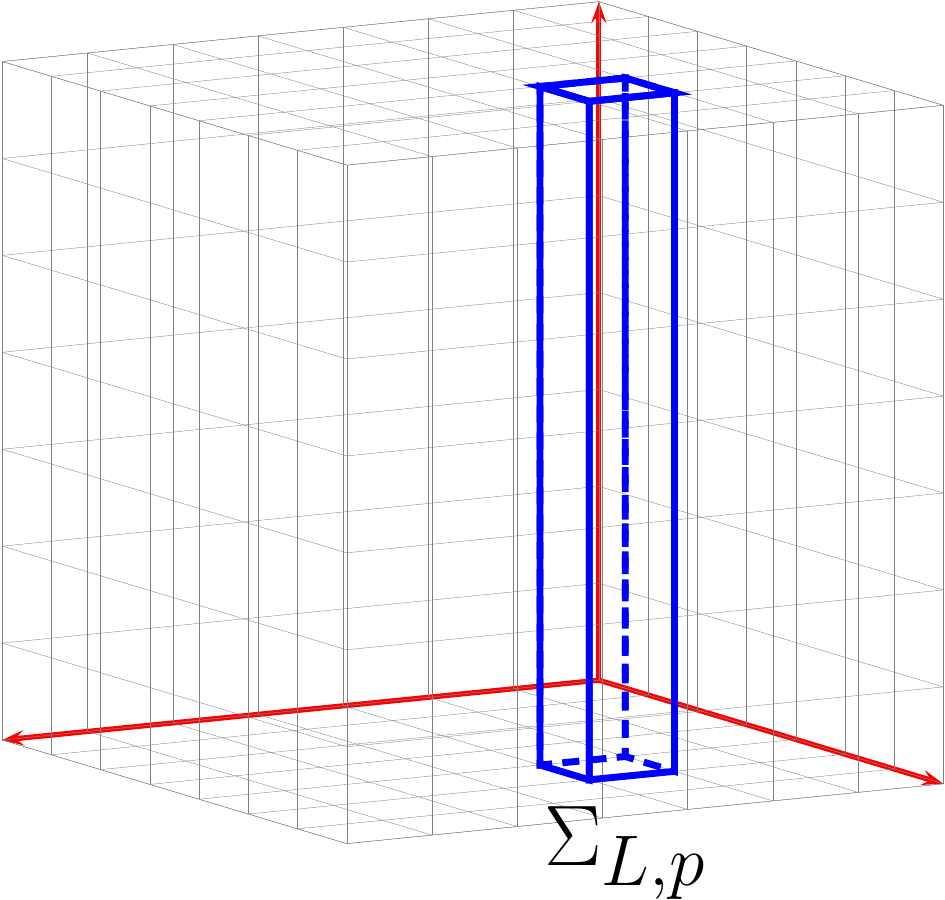}
  \end{center}
  \caption{A quasi-one-dimensional tube}
  \label{fig:6}
\end{figure}

The argument starts with decomposing the cubes $\Lambda_L$ into
quasi-one-dimensional tubes (see Figure~\ref{fig:6}) and restricting
$H_{\omega,L}^N$ to these tubes under insertion of additional Neumann
boundary conditions.
Thus let
\[\hskip-6cm A_L := \{ p\in \Z^{d-1}: -L \le p_j \le L \;\mbox{for}\;
j=1,\ldots, d-1\} \] and, for each $p\in A_L$,
\[\hskip-6cm \Sigma_{L,p} := \bigcup_{k=-L}^L \Lambda_1((p,k)),\]
with $\Lambda_1((p,k))$ denoting unit cubes centered at $(p,k)\in
\Z^d$. By $H_{\omega,L,p}^N$ we denote the restriction of $H_{\omega}$
to $L^2(\Sigma_{L,p})$ with Neumann boundary conditions. By Neumann
bracketing we have
\[ H_{\omega,L}^N \ge \bigoplus_{p\in A_L} H_{\omega,L,p}^N \] and
therefore
\begin{equation} \label{eq:tubesplitting} \min \sigma(H_{\omega,L}^N)
  \ge \min_{p\in A_L} \min \sigma(H_{\omega,L,p}^N).
\end{equation}

For a given $\omega$ such that $\omega_n \in {\mathcal C}$ for all $n$
we will say that two neighboring unit cubes are {\it matching} if the
single site potentials in the two cubes are mirror images under
reflection at the common boundary of the cubes. For each $p\in A_L$
consider the event
\[ X_{L,p} := \{ \omega: \Sigma_{L,p} \;\mbox{contains at least one
  neighboring pair of non-matching cubes}\} \] and let $X_L :=
\cap_{p\in A_L} X_{L,p}$.

By independence we have
\[ \PP(\omega \notin X_{L,p}) \le \mu^{-2L},\] where $\mu := 1/\max_{a
  \in {\mathcal C}} \mu(a)$ (after $\omega_{p,-L}$ is chosen, the
other $2L$ values of $\omega_{p,k}$ are determined by matching). This
implies that
\begin{equation} \label{eq:XLprob} \PP(X_L) \ge 1 - (2L+1)^{d-1}
  \mu^{-2L}.
\end{equation}

If $\omega \in X_L$, then each tube $\Sigma_{L,p}$ contains at least
one non-matching pair of neighboring cubes. Thus by
Proposition~\ref{prop:kn} below $\min \sigma(H_{\omega,L,p}^N) \ge E_0
+ C/L^2$ for a $C>0$ independent of $p$ and $L$. Thus
Theorem~\ref{thm6} follows from (\ref{eq:tubesplitting}) and
(\ref{eq:XLprob}).

In the proposition which was used here we can without loss consider
$p=0$:

\begin{proposition} \label{prop:kn} There is a constant $C>0$,
  independent of $L$, such that for every $\omega$ with $\omega_n\in
  {\mathcal C}$ for all $n$ and at least one non-matching pair of
  cubes in $\Sigma_{L,0}$ it holds that
  \begin{equation} \label{eq:kn} \min \sigma(H_{\omega,L,0}) \ge E_0 +
    \frac{C}{L^2}.
  \end{equation}
\end{proposition}

This, at least essentially, is a result proven in \cite{KN2}. We will
not reproduce the details of the proof, which involve a surprisingly
rich combination of analysis tools such as a Poincar{\'e}-type inequality,
the so-called ground state transform and properties of a
Dirichlet-to-Neumann operator, as well as some combinatorics. But we
will outline the main idea:

The proof of Proposition~\ref{prop:kn} starts by first arguing that is
suffices to assume that
\begin{enumerate}
\item the first two cubes $\Lambda_1((0,-L))$ and
  $\Lambda_1((0,-L+1))$ are non-matched, while
\item all other neighboring pairs are matched.
\end{enumerate}
Seeing this is not entirely trivial and requires a trick as well as
some combinatorics. The trick consists in extending the operator
$H_{\omega,L,0}^N$ by reflection to a twice longer tube and to
consider the resulting operator as an operator on the torus
$[-1/2,1/2]^{d-1} \times (\R / 2(2L+1)\Z)$ with periodic boundary
conditions. Due to symmetry this operator has the same spectral
minimum as $H_{\omega,L,0}^N$. Now one argues that the torus can be
decomposed into subsegments each of which has a non-matching pair of
cubes at one end and otherwise only matching pairs of
cubes. Introducing additional Neumann conditions on the subsegments
lowers the spectrum and it now suffices to prove the claim for each
subsegment (which has length bounded by $2(2L+1)$). Justifying that
this decomposition is possible ``is an easy combinatorics, though
somewhat lengthy to write down using symbols'', where we use the words
of \cite{KN2} and omit the details.

Under the additional assumptions (1), (2), Proposition~\ref{prop:kn}
is essentially a special case of Theorem~2.1 in \cite{KN2}. While our
situation does not satisfy the exact symmetry assumptions of
Theorem~2.1 in \cite{KN2}, the construction of our potential via
matching cubes allows to mimic the proof in \cite{KN2} almost line by
line. (We mention, however, that Section~4 of \cite{KN2} provides a
slightly different argument which allows to directly use their
Theorem~2.1 to prove Theorem~\ref{thm6} above.)

Here the results of Section~\ref{sec:uniqueness} enter as follows: As
the first pair of cubes $\Lambda_1((0,-L))$ and $\Lambda_1((0,-L+1))$
does not match, the lowest eigenvalue of the Neumann problem on the
union $\Lambda_2$ of these two cubes is strictly larger than
$E_0$. This follows from the arguments in the proof of
Theorem~\ref{thm3}, more specifically the argument at the end of
Section~\ref{sec:uniqueness} which ruled out that in this situation
the lowest eigenvalue can be equal to $E_0$. It is this operator which
plays the role of the operator $P_0^N$ in \cite{KN2}. By the results
of Section~\ref{sec:uniqueness} it is clear that inserting an
additional Neumann condition along the surface separating $\Lambda_2$
and $\Sigma_{L,0}\setminus \Lambda_2$ strictly lowers the ground state
energy of $H_{\omega,L,0}^N$ to $E_0$. The meaning of Theorem~2.1 of
\cite{KN2} is that it provides the quantitative lower bound $CL^{-2}$
on how much the energy is lowered depending on the length of the
attached tube.

The value of the constant $C>0$ in (\ref{eq:kn}) which is provided by
the argument in \cite{KN2} will depend on the choice of
$\omega_{0,-L}$ and $\omega_{0,-L+1}$, after which the remaining
values of $\omega_n$ are determined by matching. However, as only the
finitely many values in the corners ${\mathcal C}$ are allowed, one
can ultimately choose the smallest of finitely many values of $C$.

\vspace{.3cm}

As indicated earlier, Theorem~\ref{thm6} is the result which will be
used later, rather than its consequences for the IDS $N(E)$ of
$H_{\omega}$. However, we mention that the following Lifshitz tail
bound can be derived from Theorem~\ref{thm6} with standard arguments,
see \cite{KN2}:
\begin{equation} \label{eq:Lifshitz} \limsup_{E\downarrow E_0}
  \frac{\log |\log N(E)|}{\log (E-E_0)} \le -\frac{1}{2}.
\end{equation}

The Lifshitz exponent $-1/2$ obtained here is likely not optimal. One
would expect that the correct exponent is $-d/2$, as known for the
Anderson model or Poisson model. The reason for the discrepancy lies
in the essentially one-dimensional argument which enters the proof
through the decomposition of cubes into quasi-one-dimensional tubes.

We stress the fundamentally different low energy behavior of the IDS
of the RDM in the one-dimensional and multi-dimensional settings. If
supp$\,\mu = {\mathcal C}$ with equal probability for all corners, then
(\ref{eq:Lifshitz}) shows that the IDS has a very thin tail for $d\ge
2$, while by Theorem~\ref{thm5} it has a very fat tail (and thus the
spectrum does {\it not} have a fluctuation boundary) for $d=1$. From
this point of view it is a fortunate coincidence that our main goal
here is to prove localization in $d\ge 2$ and that, as discussed in
the Introduction, localization for the one-dimensional case was
already settled by very different methods.

\section{The Missing Link} \label{sec:missinglink}

It is tempting to believe at this point of our work that we are
halfway done with verifying the necessary ingredients for a multiscale
analysis proof of localization for the RDM. Under suitable assumptions
we have shown the Lifshitz-tail bound (\ref{eq:Lifshitzbound}), so it
remains to establish a Wegner estimate. Unfortunately, the assumption
that supp$\,\mu$ be finite in Theorem~\ref{thm6} makes us face a
dilemma: Most known proofs of Wegner-type estimates, with the
exception of some results in $d=1$ \cite{CKM, DSS}, require some
smoothness or at least continuity of the distribution of the random
parameters, due to the use of averaging techniques involving only
finitely many random parameters. For the multi-dimensional continuum
Anderson model with Bernoulli distributed random coupling constants a
localization proof near the bottom of the spectrum was enabled only
recently by the powerful extension of multiscale analysis presented in
\cite{BK}, see also \cite{AGKW} for an extension to the case of
arbitrary single site distributions and \cite{GK11} for a detailed
elaboration of the intricate ideas behind \cite{BK}. One of the main
features of this approach is that the Wegner estimate is not
established as an a-priori-ingredient, but its proof is part of the
multiscale iteration procedure leading to localization. We mention
that due to the use of unique continuation arguments this approach
does not work on the lattice, leaving the proof of localization for
the multi-dimensional discrete Bernoulli-Anderson model an open
problem.

Thus, if we want to complete a localization proof for the RDM based on
``traditional'' multiscale analysis, the proof of a Wegner estimate
will likely require a sufficient amount of regularity of the
distribution $\mu$ of the displacement parameters $\omega_n$. But this
means that we also need to extend the Lifshitz-tail bound to more
general distributions $\mu$. The proof discussed in
Section~\ref{sec:SpecLifshitz} above does not extend to this case, as
it would require to take an infimum over infinitely many positive
constants $C$ with insufficient quantitative information available to
guarantee that the resulting constant is strictly positive.

It turned out that the missing link which allowed to overcome both
remaining problems, the extension of the Lifshitz-tail bound to a
larger class of distributions and the proof of a Wegner estimate for
this class, is provided by an inconspicuous but crucial improvement on
how bubbles tend to the corners, meaning Theorem~\ref{thm2}. There it
was shown that the function $E_0(a)$, as long as it does not vanish
identically, is strictly decreasing in each of its variables away from
the origin. The crucial improvement is that this decrease arises in
the form of {\it non-vanishing derivative}.

This is the first instance where we will have to require some
smoothness of $q$, as differentiability of $E_0(\cdot)$ requires
differentiability of $q$ via perturbation theory, see Section~2.1 of
\cite{KLNS}. For convenience, we will assume that $q$ is $C^{\infty}$,
even if much less is needed below and in the rest of this paper.

As this is the last time we add assumptions on $q$, let us restate the
full set:

\vspace{.3cm}

{\bf (A1)'} The single-site potential $q:\R^d \to \R$ is infinitely
differentiable, reflection-symmetric in each variable and such
supp$\,q \subset [-r,r]^d$ for some $r<1/4$. Also assume that $E_0(a)
= \min \sigma(H_{\Lambda_1}^N(a))$ does not vanish identically in
$a\in \overline{G}$.

\vspace{.3cm}

We now get

\begin{theorem} \label{thm7} Assume {\bf (A1)'}. Then for all
  $a=(a_1,\ldots,a_d) \in \overline{G}$ and all $i=1,\ldots,d$ we have
  \[ \partial_i E_0(a) \left\{ \begin{array}{ll} <0, & \mbox{if
        $a_i>0$}, \\ = 0, & \mbox{if $a_i=0$}, \\ >0, & \mbox{if
        $a_i<0$}. \end{array} \right. \]
\end{theorem}

The proof of Theorem~\ref{thm7} is far from obvious (at least to us)
and is best discussed in the larger context of considering similar
questions for more general domains $G$. In order to not interrupt the
presentation of our main story, i.e.\ the proof of localization for
the random displacement model, we postpone this discussion to
Section~\ref{sec:bubblesresults} below. But let us point out that the
step from Theorem~\ref{thm2} to Theorem~\ref{thm7} turned out to be
far from straightforward. The ``smooth methods'' behind the proof of
Theorem~\ref{thm7} are very different from the symmetry-based operator
theoretic methods used to prove Theorem~\ref{thm2} in \cite{BLS1} and,
in particular, explicitly use second order perturbation theory.

\section{General Lifshitz tails} \label{sec:GenLifshitz}

The first of two important applications of Theorem~\ref{thm7} is that
it allows us to extend the Lifshitz tail bound found in
Theorem~\ref{thm6} to general distributions $\mu$, not requiring
finiteness of the support.

\begin{theorem} \label{thm8} Assume that $q$ and $\mu$ satisfy {\bf
    (A1)'} and {\bf (A2)}. Then there exist $C_1>0$ and $\mu>1$ such
  that
  \begin{equation} \label{eq:Lifshitzbound2} \PP \left( \min
      \sigma(H_{\omega,L}^N) < E_0 + \frac{C_1}{L^2} \right) \le
    (2L+1)^{d-1} \mu^{-2L}
  \end{equation}
  for all $L\in \N$.
\end{theorem}

We will prove this result by comparing the quadratic form of
$H_{\omega,L}^N$ with the quadratic form of a modified displacement
model where all bubbles have been moved to the closest corner within
their cell. Thus, for $a\in \overline{G}$, let $c(a) \in {\mathcal C}$
be the corner closest to $a$ (if several corners are equally close,
any of them can be chosen). For a displacement configuration $\omega =
(\omega_n)_{n\in \Z^d} \in \overline{G}^{\Z^d}$, define $c(\omega) \in
\overline{G}^{\Z^d}$ by $(c(\omega))_n = c(\omega_n)$.

From Theorem~\ref{thm2} we know that the single-site operator
$H_{\Lambda_1}^N(c(a))$ has lower ground state energy than
$H_{\Lambda_1}^N(a)$. Theorem~\ref{thm7} allows us to quantify this,
saying that the distance of the two ground state energies is
proportional to $|a-c(a)|$. In particular, there exists $C_2 \in
(0,\infty)$ such that
\begin{equation} \label{eq:linear} E_0(a)-E_0 \ge \frac{1}{C_2} D(a),
\end{equation}
where $D(a) = \min_{c\in {\mathcal C}} |a-c|$. This is one of the two
central ingredients in the proof of the following result. The other
one will be Neumann bracketing.

\begin{proposition} \label{prop:formcompare} There exists a constant
  $C_3 \in (0,\infty)$ such that, in the sense of quadratic forms,
  \begin{equation} \label{eq:formcompare} H_{\omega,L}^N - E_0 \ge
    \frac{1}{C_3} (H_{c(\omega),L}^N-E_0).
  \end{equation}
  for all $\omega \in \overline{G}^{\Z^d}$ and all $L\ge 0$.
\end{proposition}

In particular, (\ref{eq:formcompare}) implies
\begin{equation} \label{eq:energycompare} \min \sigma(H_{\omega,L}^N)
  -E_0 \ge \frac{1}{C_3} ( \min \sigma(H_{c(\omega),L}^N -E_0).
\end{equation}
The RDM $H_{c(\omega)}$ has i.i.d.\ distributed displacements supported
on $\mathcal C$ and thus satisfies the assumptions of
Theorem~\ref{thm6}. Therefore, with $C$ and $\mu$ from
Theorem~\ref{thm6},
\[ \PP \left( \min \sigma(H_{\omega,L}^N) -E_0 < \frac{C}{C_3 L^2}
\right) \le \PP \left( \min \sigma(H_{c(\omega),L}^N ) - E_0 <
  \frac{C}{L^2} \right) \le (2L+1)^{d-1} \mu^{-2L}, \] proving
Theorem~\ref{thm8}.

Thus it remains to prove Proposition~\ref{prop:formcompare}. The
strategy for this is to first prove a corresponding result for the
single-site operators $H_{\Lambda_1}^N(a)$ and then extend this by
Neumann bracketing to the operators $H_{\omega,L}^N$. For the
single-site operators one separately considers the cases where $a$ is
close to a corner or not close to a corner.

\begin{lemma} \label{lem:close} There exist $C>0$ and $\delta>0$ such
  that, if $D(a) \le \delta$, then
  \begin{equation} \label{eq:close} H_{\Lambda_1}^N(a) - E_0 \ge
    \frac{1}{C} (H_{\Lambda_1}^N(c) -E_0 + |a-c|).
  \end{equation}
\end{lemma}

\begin{lemma} \label{lem:far} Fix $\delta \in (0,1)$. There exists
  $C_{\delta} \in (0,\infty)$ such that, for $D(a) \ge \delta$ and all
  $c\in {\mathcal C}$,
  \begin{equation} \label{eq:far} H_{\Lambda_1}^N(a) - E_0 \ge
    \frac{1}{C_{\delta}} (H_{\Lambda_1}^N(c) -E_0 + |a-c|).
  \end{equation}
\end{lemma}

Before discussing the proofs of the two Lemmas, let us show how we use
them to prove Proposition~\ref{prop:formcompare}. Note that, applying
Lemma~\ref{lem:far} with $\delta$ as provided in
Lemma~\ref{lem:close}, both Lemmas combined prove the $L=0$ case. To
extend this to general boxes we employ an argument previously used in
the proof of Theorem~2.1 in \cite{KN1}. It is crucial here that we
work with Neumann boundary conditions.

For $\psi \in H^1(\Lambda_{2L+1})$, the form domain of
$H_{\omega,L}^N$, one has that the restriction of $\psi$ to
$\Lambda_1(n)$ is in $H^1(\Lambda_1(n))$ for each $n\in
\Lambda_{2L+1}' := \Lambda_{2L+1} \cap \Z^d$. Moreover,
\[ \langle (H_{\omega,L}^N - E_0) \psi, \psi \rangle = \sum_{n\in
  \Lambda_{2L+1}'} \langle (H_{\Lambda_1(n)}^N(\omega_n) -E_0)\psi,
\psi \rangle, \] where we work with the usual slightly abusive
notation for quadratic forms.

The same argument may be applied to $H_{c(\omega)}$,
\[ \langle (H_{c(\omega),L}^N - E_0) \psi, \psi \rangle = \sum_{n\in
  \Lambda_{2L+1}'} \langle (H_{\Lambda_1(n)}^N(c(\omega_n)) -E_0)
\psi, \psi \rangle. \] Now Proposition~\ref{prop:formcompare} follows
by applying Lemmas~\ref{lem:close} and \ref{lem:far} for each $n$,
summing, and omitting the positive term $\sum_n \langle |\omega_n -
c(\omega_n)| \psi, \psi \rangle_{\Lambda_1(n)}$.

Before we can end this section, we still owe a discussion of the
proofs of Lemmas~\ref{lem:close} and \ref{lem:far}. To see the latter,
note that $D(a) \ge \delta$ implies $H_{\Lambda_1}^N(a)-E_0 \ge
\delta/C_2$ by (\ref{eq:linear}). Using the rough bound $|q(x-a) -
q(x-c)| \le 2\|q\|_{\infty}$ and setting $C:= 1 +
2C_2\|q\|_{\infty}/\delta$ we get
\[ (C+1) (H_{\Lambda_1}^N(a) - E_0) - (H_{\Lambda_1}^N(c) -E_0) \ge
\frac{C\delta}{C_2} - 2\|q\|_{\infty} = \frac{\delta}{C_2},\] and thus
\[ H_{\Lambda_1}^N(a) - E_0 \ge \frac{1}{C+1} \left(
  H_{\Lambda_1}^N(a) - E_0 +\frac{\delta}{C_2} \right).\] As $\delta =
\frac{\delta}{|a-c|} |a-c| \ge \frac{\delta}{2d_{max}\sqrt{d}} |a-c|$,
(\ref{eq:far}) follows with $1/C_{\delta}$ chosen as the smaller of
$1/(C+1)$ and $\delta/(2(C+1)C_2 d_{max} \sqrt{d})$.

The previous argument doesn't use the full strength of
(\ref{eq:linear}), but only that $E_0(\cdot)$ is continuous and
strictly minimized in the corners. The proof of Lemma~\ref{lem:close}
is more subtle and depends on the linear growth of $E_0(\cdot)$ away
from the corners. To $a\in \overline{G}$ pick $c\in {\mathcal C}$ such
that $D(a)=|a-c|$. By smoothness of $q$ we have the Taylor
approximation $q(\cdot -a) - q(\cdot -c) = (c-a) \cdot \nabla q(\cdot -c) +o(|a-c|)$ and
thus
\begin{equation} \label{eq:taylor} H_{\Lambda_1}^N(c) -E_0 =
  H_{\Lambda_1}^N(c) - E_0 + (c-a) \cdot \nabla q(\cdot -c) + o(|a-c|).
\end{equation}
Bounding the left hand side by (\ref{eq:linear}) we get, in the sense
of quadratic forms,
\[ H_{\Lambda_1}^N(c) - E_0 + (c-a) \cdot \nabla q(\cdot -c) \ge \frac{1}{C_2}
|a-c| + o(|a-c|).\] Hence, for $\rho \in (0,1)$ sufficiently small and
$\sigma \in {\mathcal S}^{d-1}$ with $a=c+\rho \sigma \in
\overline{G}$,
\[ H_{\Lambda_1}^N(c) - E_0 - \rho \sigma \cdot \nabla q(\cdot -c) \ge
\frac{\rho}{2C_2}.\] We apply Lemma~\ref{lem:aux} below with $A=
H_{\Lambda_1}^N(c)-E_0$ and $B= -\rho \sigma \cdot \nabla q(\cdot-c)$ to
conclude that for $C_{\rho} = \max(2,2C_2/\rho)$ $t\in [0,1/2]$ and
$\sigma \in {\mathcal S}^{d-1}$ with $c+\rho \sigma \in \overline{G}$,
\[ H_{\Lambda_1}^N(c) -E_0 -t \rho \sigma \cdot \nabla q(\cdot -c) \ge
\frac{1}{C_{\rho}} (H_{\Lambda_1}^N(c) - E_0 +t).\] From this and
(\ref{eq:taylor}) we find for $|a-c|\le \rho/2$ and $t=|a-c|/\rho$,
\[ H_{\Lambda_1}^N(a)-E_0 \ge \frac{1}{C_{\rho}}
(H_{\Lambda_1}^N(c)-E_0 +|a-c|/\rho) + o(|a-c|).\] This implies
(\ref{eq:close}) if $\delta>0$ is chosen sufficiently small,
completing the proof of Lemma~\ref{lem:close}.

We have used the following simple fact, which was previously used in a
similar context in \cite{KN1}.
\begin{lemma} \label{lem:aux} Let $A$ be self-adjoint and $B$ bounded
  and self-adjoint with $A\ge 0$ and $A+B\ge c_0 >0$, then
  \[ A+tB \ge \min(\frac{1}{2}, c_0) \cdot (A+t) \] for all $t\in [0,1/2]$.
\end{lemma}

This is elementary:
\[ A+tB = (1-t)A + t(A+B) \ge \frac{1}{2} A +tc_0 \ge
\min(\frac{1}{2}, c_0) (A+t).\]

\section{Wegner Estimate} \label{sec:Wegner}

To describe the ideas behind the proof of a Wegner estimate, we consider $H^r_\omega$, a suitable random Schr{\"o}dinger operator on $L^2(\R^d)$. Let $L>0$ and  $H^r_{\omega,L}$ be the restriction of $H^r_{\omega}$ to the cube $\Lambda_L$ with, say,
Dirichlet boundary conditions. The boundary conditions are expected
not to play a too important role.

A {\it Wegner estimate} (see~\cite{We:81}) is an estimate on
\begin{equation} \label{Wegnerquant}
  \esp(\tr \chi_{[E_0-\varepsilon,E_0+\varepsilon]}(H^r_{\omega,L}))
  =\esp({\#\{\text{eigenvalues of } H^r_{\omega,L}\text{ in }
    [E_0-\varepsilon,E_0+\varepsilon] \}})
\end{equation}
for $L$ large, $\varepsilon$ small and a fixed energy $E_0$. It
can also take the form of an estimate on the probability
$\pro\{H^r_{\omega,L}\text{ has an eigenvalue in }
[E_0-\varepsilon,E_0+\varepsilon]\}$ which, by Chebyshev's inequality, is
smaller than the previous quantity.

From their very form, it is clear that both quantities should increase
with $\varepsilon$ and with $L$. The existence of an integrated
density of states for $H_\omega^r$ suggests that the optimal upper
bound should be proportional to $|\Lambda_L| \sim L^d$. The optimal upper bound in
$\varepsilon$ is related to the regularity of the integrated density
of states. The best bound one may expect is of the form
$C\varepsilon L^d$.

Let us give the heuristic underlying such a bound in the simplest
case, the case when $H^r_\omega$ is the continuous Anderson type model
$H^A_{\lambda(\omega)}$ defined in~\eqref{eq:Anderson}, when $q$ has a
fixed sign, say, positive, is continuous and bounded, its support
contains $\Lambda_1$ and the coupling constants
$(\lambda_n)_{n\in\Z^d}$ are i.i.d.\ and bounded.

Let $(E_j(\omega,L))_j$ denote the eigenvalues of $H^A_{\lambda(\omega),L}$
ordered increasingly. To estimate the quantity
$\esp(\tr \chi_{[E_0-\varepsilon,E_0+\varepsilon]}(H^A_{\lambda(\omega),L}))$,
we can write
\begin{align*}
  \esp(\tr \chi_{[E_0-\varepsilon,E_0+\varepsilon]}(H^A_{\omega,L}))&
  =\esp \left(\sum_{j}
    \chi_{[E_0-\varepsilon,E_0+\varepsilon]}(E_j(\omega,L)) \right) \\
  &\leq \sum_{j\in N_L}\pro\left\{
    E_j(\omega,L)\in [E_0-\varepsilon,E_0+\varepsilon] \right\}.
\end{align*}
where, by standard bounds on Schr{\"o}dinger operators, $\#N_L\lesssim
L^d$.

In the case of the continuous Anderson model under the assumptions
made above, for $\alpha>0$, the operator inequality $
H^A_{\lambda(\omega)+\overline{\alpha}}-H^A_{\lambda(\omega)}
\gtrsim\alpha$ tells us that
\begin{equation}
  \label{eq:1}
  \forall j,\quad E_j(\lambda(\omega)+\overline{\alpha},L)-
  E_j(\lambda(\omega),L)\gtrsim \alpha,
\end{equation}
where $\overline{\alpha}$ is the vector whose entries are all
$\alpha$.

Based on this and under the assumption that the distribution of the $\lambda_n$ has a bounded density one can prove that
\begin{equation}
  \label{linbound}
  \forall j,\quad \pro\left\{E_j(\omega,L)\in[E_0-\varepsilon,
    E_0+\varepsilon] \right\}\lesssim\varepsilon
\end{equation}
and obtains the desired bound in $\varepsilon L^d$. The proof of (\ref{linbound}), while essentially based on the ideas described above, requires additional technical work.

For the discrete $d$-dimensional Anderson model, the bound
\begin{equation} \label{linWegner}
\esp(\tr \chi_{[E_0-\varepsilon,E_0+\varepsilon]}(H^r_{\omega,L})) \le C \varepsilon L^d
\end{equation}
essentially goes back to Wegner's original paper \cite{We:81}, with some technical details filled in later. A detailed proof, essentially following Wegner's orginal argument, can be found, for example, in the recent survey \cite{Kirsch2}.
Obtaining the bound (\ref{linWegner}) for the continuum Anderson model was harder, as one can not use the same rank one perturbation methods as in the discrete case to control the spectral shift due to single site terms. Initially, a bound of the form $C \varepsilon L^{2d}$ was obtained for the continuum Anderson model in \cite{Kirsch96} (for a proof with slightly different methods see also \cite{Stollmann}). For $q$ of fixed sign, but without the assumption that the support of $q$ contains $\Lambda_1$ and at arbitrary energy, the linear in volume bound (\ref{linWegner}) was ultimately obtained in \cite{CHK}.

To describe how a Wegner estimate for the random displacement model considered here was found, let us describe a generalization of the idea outlined above which goes back to \cite{Klopp,Klopp95,MR1684138}. To estimate $\pro\{E_j(\omega,L)\in
[E_0-\varepsilon,E_0+\varepsilon]\}$, we study the mapping
$\omega\mapsto E_j(\omega,L)$ that realizes a
``projection'' from the parameter (probability) space onto the real axis; and we want
to measure the size (with respect to the probability measure on the
parameter space) of the pre-image of some interval.
The idea is then
to find a vector field ${\mathcal V}$ in the parameters $\omega$ such
that the eigenvalue $E_j(\omega,L)$ moves when $\omega$ moves along
the flow of the vector field.
\begin{figure}[h]
   \centering
   \includegraphics[bbllx=2124,bblly=2052,bburx=2585,bbury=2444,height=6cm]{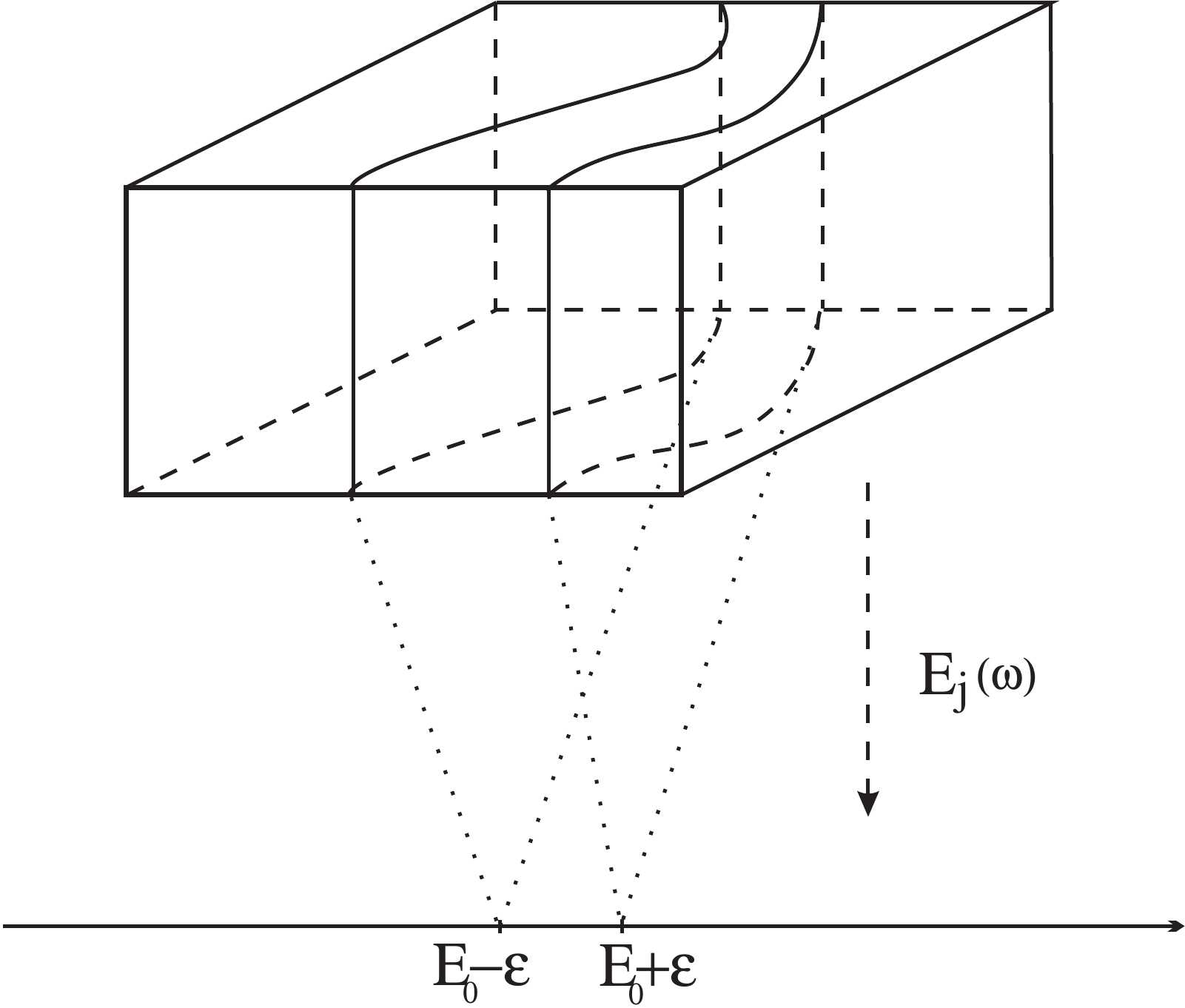}
   \caption{Foliation and projection of the probability space}
    \label{fig:7}
 \end{figure}
%
%\begin{figure}[h] \label{fig:7}
%   \centering
%   \includegraphics[bbllx=2124,bblly=2052,bburx=2585,bbury=2444,height=6cm]{des-viet3.pdf}
%   \caption{Foliation and projection of the probability space}
% \end{figure}
%
The flow of ${\mathcal V}$ foliates the
parameter space nicely and the volume we want to measure is just the
volume contained in a layer between two leaves (see
Figure~\ref{fig:7}). This volume will then be of size the width of
this layer at least when the probability measure has a regular
density. So if one is able to do this for all the eigenvalues, one
gets an estimate of the form $\varepsilon L^d$.

To be able to do this for all eigenvalues at a time, one may choose
${\mathcal V}$ so that $H_\omega^r$ differentiated along ${\mathcal
  V}$ has nice properties (e.g.\ positivity). Let us take the simple
example of the continuous Anderson Hamiltonian under the assumptions
made above. If we take $\displaystyle{\mathcal
  V}=\text{div}_{\lambda(\omega)}=\sum_{n\in \Lambda_L}
\frac\partial{\partial\lambda_n(\omega)}$, then ${\mathcal
  V}H_{\lambda(\omega)}^L\gtrsim \chi_{\Lambda_L}$. This ensures
${\mathcal V}E_j(\omega,L)\gtrsim 1$ which is
exactly~\eqref{eq:1}. This is what is needed for a Wegner estimate for
the continuous Anderson Hamiltonian when the single site coupling
constants admit a bounded density.
\par The right choice of vector field is model dependent; for the
Anderson model, discrete or continuous, in many cases, one may use the
divergence vector field as above (for example,
see~\cite{We:81,Stollmann}). Another useful vector field with respect
to this problem is the generator of the dilations $\sum_n
\lambda_n(\omega)\partial_{\lambda_n(\omega)}$. It can be used to get
a Wegner estimate for the continuous Anderson model without sign
assumptions on $V$ \cite{Klopp95,HislopKlopp} and, in certain cases,
for the random displacement models \cite{Ghribi/Klopp}. Other types of
randomness may require different types of vector fields, see
e.g.~\cite{Klopp,2010arXiv1012.5185E}.

\vspace{.5cm}

For the random displacement Hamiltonian $H_\omega$, we will use the
same idea and introduce a new vector field.  The choice of this vector
field is motivated by Theorems~\ref{thm2} and \ref{thm7}, which
indicate that at least low lying eigenvalues should decrease
monotonically if the single site potentials are moved towards a corner
of their cell.

For a function $f$ on $G$ we set
\[
(\partial_c f)(a) := \frac{c(a)-a}{|c(a)-a|} \cdot \nabla f(a),
\]
with $c(a)$ denoting the corner closest to $a$ as in
Section~\ref{sec:GenLifshitz}. Thus, $\partial_c$ denotes the
directional derivative in the direction of the closest corner, where
points $a$ with multiple closest corners will not play a role in the
arguments below (starting from (\ref{eq:cutoff}) below we introduce a
cut-off which restricts the values of $a$ relevant for the proof to
small neighborhoods of the corners).

By Theorem~\ref{thm7} there exist $\delta_0>0$ and $r_0>0$ such that
\begin{equation}
  \label{eq:boundonderiv}
  \partial_{c}E_0(a) \leq -\delta_0 \quad \mbox{for all $a\in A_{r_0}
    := \{a \in G: |c(a)-a| \leq r_0\}$},
\end{equation}
a neighborhood of $\mathcal{C}$.

Let $\eta\in C^\infty(\mathbb{R})$ such that $0\le \eta \le 1$,
$\eta(r)=1$ for $r\leq r_0$ and $\eta(r)=0$ for $r \geq 2r_0$. Using
this function as a cut-off, we localize the vector fields associated
with $\partial_c$ onto a neighborhood of the corners, defining
\begin{equation} \label{eq:cutoff} (\partial'_c f)(a) :=
  \eta(|c(a)-a|) (\partial_c f)(a).
\end{equation}

For each $n\in \mathbb{Z}^d$, we write
\begin{equation} \label{eq:Hamderiv}
  \partial'_{c,\omega_n}H_\omega
  =\partial'_{c,\omega_n} q(\cdot-n-\omega_n)
  =-\eta(|c(\omega_n)-\omega_n|)
  \frac{c(\omega_n)-\omega_n}{|c(\omega_n)-\omega_n|}  \cdot(\nabla q)(\cdot -n-\omega_n).
\end{equation}

If $\psi \in H^1(\Lambda_{2L+1})$, the form domain of $H^N_{\omega,L}$,
then $\psi_n := \psi|_{\Lambda_1(n)} \in H^1(\Lambda_1(n))$, the form
domain of $H_n(\omega_n)$, and, with the usual abuse of notation for
the quadratic form,
\begin{equation} \label{eq:formdecomp} \langle \psi, H^N_{\omega,L} \psi
  \rangle = \sum_{n\in \Lambda_{2L+1}'} \langle \psi_n, H_n(\omega_n)
  \psi_n \rangle,
\end{equation}
as well as
\begin{equation} \label{eq:derlocal} \sum_{n\in \Lambda_{2L+1}'}
  \langle \psi, \partial'_{c,\omega_n} H^N_{\omega,L} \psi \rangle =
  \sum_{n\in \Lambda_{2L+1}'} \langle \psi_n, \partial'_{c,\omega_n}
  H_n(\omega_n) \psi_n \rangle.
\end{equation}

\begin{proposition} \label{prop:key} There exist $\delta_1>0$ and
  $\delta_2>0$ such that
  \begin{equation} \label{eq:key} -\sum_{n\in \Lambda_{2L+1}'} \langle
    \psi, (\partial'_{c,\omega_n} H^N_{\omega,L}) \psi \rangle \ge
    \delta_1 \|\psi\|^2
  \end{equation}
  for all $L\in \N$, and $\psi \in H^1(\Lambda_{2L+1})$ with $\langle
  \psi, (H^N_{\omega,L}-E_0)\psi \rangle \le \delta_2 \|\psi\|^2$.
\end{proposition}
\noindent Thus, near the bottom of the spectrum of $H_{\omega,L}$, the
vector field $\displaystyle-\sum_{n\in
  \Lambda_{2L+1}'}\partial'_{c,\omega_n}$ satisfies exactly the
property we are looking for to proceed according to the heuristics
explained above. For the details of the proof of Proposition~\ref{prop:key} as well as for the proof of the implied Wegner estimate in Theorem~\ref{thm9} below we refer to \cite{KLNS}.

In order to exploit Proposition~\ref{prop:key} we formulate the following
final set of assumptions for the distribution $\mu$ of the i.i.d.\
random displacement parameters $\omega = (\omega_n)$:

\vspace{.3cm}

{\bf (A2)'} With $G$ and $\mathcal C$ as above, let ${\mathcal C}
\subset \mbox{supp}\,\mu \subset \overline{G}$. Also assume that there
exists a neighborhood of $\mathcal C$ on which $\mu$ has a
$C^1$-density.

\vspace{.3cm}

More formally, this means that there exists $\varepsilon>0$ and a
$C^1$-function $\rho: \overline{G} \to \R$, such that for every $S
\subset \cup_{a\in {\mathcal C}} \{ x: |x-a|<\varepsilon\} \cap
\overline{G}$ we have
\[ \mu(S) = \int_S \rho(x)\,dx.\]

This guarantees that the random variables $|c(\omega_n)-\omega_n|$, which parametrize the layers of the vector field, have absolutely continuous distribution near $0$ (i.e.\ for values of $\omega_n$ near the corners). Thus we have exactly the situation required by the above heuristics, which indeed leads to the following Wegner estimate:

\begin{theorem} \label{thm9} Assume {\bf (A1)'} and {\bf (A2)'}. Then
  there exists $\delta>0$ such that, for any $\alpha \in (0,1)$, there
  exists $C_{\alpha}>0$ such that, for every interval $I \subset [E_0,
  E_0+\delta]$ and $L\in \N$,
  \begin{equation} \label{eq:Wegner} \E( \mbox{tr}\,
    \chi_I(H_{\omega,L}^N)) \le C_{\alpha} |I|^{\alpha} L^d.
  \end{equation}
\end{theorem}

We finally need to comment on the reason for the appearance of the exponent $\alpha \in(0,1)$ in (\ref{eq:Wegner}). This is a final price we pay for the non-monotonicity (as well as non-analyticity) of our model in the random parameters. The latter necessitate some additional changes in the original strategy of Wegner. The proof of Theorem~\ref{thm9} in \cite{KLNS} uses an adaptation of a Wegner estimate proof developed in \cite{HislopKlopp} to handle Anderson models with sign-indefinite single-site potentials. Their argument is based on $L^p$-bounds for Krein's spectral shift function proven in \cite{CHN}. These bounds only hold for $p<\infty$ and not for $p=\infty$, which is the reason that we can't choose $\alpha=1$ in Theorem~\ref{thm9}.

\section{Localization} \label{sec:localization}

At this point we have essentially reached the end of the story which
was to be told here. As explained in the introduction, with the
Lifshitz tail bound of Theorem~\ref{thm8} and the Wegner estimate of
Theorem~\ref{thm9} available, localization for the RDM can be proven
via multi-scale analysis. The MSA method is as powerful as the details
of carrying it out are intricate. Presenting these details is not a
goal of this article. Very good introductions into the mathematics of
MSA can be found in the book \cite{Stollmann} and the surveys
\cite{Klein} and \cite{Kirsch2}, which also provide extensive
bibliographies.

The main task left to us is to state state the exact result on
localization for the RDM which was obtained in \cite{KLNS}. Here
$\chi_x$ denotes the characteristic function of a unite cube centered
at $x$, $\chi_I(H)$ the spectral projection onto $I$ for the operator
$H$, and $\|\cdot\|_2$ the Hilbert-Schmidt norm.

\begin{theorem} \label{thm10} Assume {\bf (A1)'} and {\bf (A2)'}. Then
  there exists $\delta >0$ such that $H_{\omega}$ almost surely has
  pure point spectrum in $[E_0, E_0+\delta]$ with exponentially
  decaying eigenfunctions.

  Moreover, $H_{\omega}$ is dynamically localized in $I$, in the sense
  that for every $\zeta <1$, there exists $C<\infty$ such that
  \begin{equation} \label{eq:dynloc} \E \left( \sup_{|g|\le 1} \|
      \chi_x g(H_{\omega}) \chi_I(H_{\omega}) \chi_y\|_2^2 \right) \le
    C e^{-|x-y|^{\zeta}}
  \end{equation}
  for all $x,y \in \Z^d$. The supremum is taken over all Borel
  functions $g:\R \to \C$ with satisfy $|g|\le 1$ pointwise.
\end{theorem}

We note that dynamical localization in the physical sense is covered
by (\ref{eq:dynloc}) in choosing $g(H) = e^{-itH}$ and taking the
supremum over $t\in \R$.

The subexponential decay in $|x-y|$ found in (\ref{eq:dynloc}) is the
strongest type of dynamical localization which has been obtained
through MSA. This is a result of Germinet and Klein in
\cite{Germinet/Klein}, who used a four times bootstrapped version of
the MSA argument, allowing to conclude strong forms of localization
from rather weak forms of initial length estimates provided by the
Lifshitz tail bound. As described in some more detail in \cite{KLNS},
the survey paper \cite{Klein} provides a very useful resource in
explicitly singling out all the properties of a model, which go into
the argument in \cite{Germinet/Klein}. In addition to the crucial
Lifshitz-tail and Wegner bounds proven in Theorems~\ref{thm8} and
\ref{thm9}, the RDM has all other required properties.

In this context we also recommend the book \cite{Stollmann} as a very
readable account of MSA. Similar to \cite{Klein} it clearly exhibits
the properties of a model which are needed to prove localization via
MSA. It uses a version of MSA less sophisticated than what is done in
\cite{Germinet/Klein}, essentially bootstrapping the MSA scheme just
twice, to conclude spectral localization and a weaker form of
dynamical localization, that is
\begin{equation} \label{eq:weakdynloc} \E \left( \sup_{|g|\le 1}
    \||X|^p g(H_{\omega}) \chi_I(H_{\omega}) \chi_0\| \right) < \infty
\end{equation}
for all $p>0$ in a $p$-dependent neighborhood $I$ of $E_0$, giving
power-decay rather than subexponential decay in
(\ref{eq:dynloc}). Here $|X|$ is the multiplication operator by the
length of the variable $x\in \R^d$.

\section{Bubbles Tend to the Corner} \label{sec:bubblesresults}

\subsection{Bubbles tend to the boundary} The Lifshitz tail estimate
as well as the Wegner estimate have as a basic input an estimate on
the ground state energy of a Neumann problem as a function of the
position of the potential.

It is convenient to take a more general point of view for this problem
in which the unit cell $\Lambda_1$ is replaced by a bounded domain $D$
with smooth boundary, and to consider the non-degenerate lowest
eigenvalue $E_0(a)$ of the operator
\begin{equation} \label{eq:genNeumann} -\Delta_N +q(x-a) \ .
\end{equation}
Here $\Delta_N$ is the Laplace operator with Neumann boundary
conditions on $\partial D$. We shall assume that the potential $q$ is
smooth and has compact support such that the set
$$
G=\{a \in \R^d: \supp q(\cdot -a) \subset D\}
$$
is not empty. Thus, $G$ is an open and bounded set. We shall, in
addition, assume that it is connected.

The following second order perturbation theory result sets the stage
for our investigation.  Denote by $\partial_a = w \cdot \nabla_a$ and
likewise $\partial_x = w \cdot \nabla_x$ where $w$ is a fixed vector and
the subscript denotes the variable in which we differentiate.

\begin{lemma}[\bf{Second order perturbation
    theory}] \label{perturbation} The lowest eigenvalue as a function
  of $a$ satisfies the equation
  \begin{equation} \label{eq:bigone}
    \partial_a^2 E_0 - 4 \partial_a E_0 \langle u_0, \partial_x u_0\rangle = 2 \int_D \nabla \cdot (\partial_x u_0 \nabla \partial_x u_0) dx
    - 2 \sum_{k\not= 0} \frac{B(u_k, \partial_x u_0)^2}{E_k-E_0} \ .
  \end{equation}
  Here $B(u,v)$ denotes the bilinear form
$$
B(u,v) = ( u, \Delta v)-(\Delta u, v) \ ,
$$
and $u_k(x;a)$ is the eigenfunction associated with the eigenvalue
$E_k(a), k=0,1,2, \dots$.
\end{lemma}
The proof of this lemma can be found in \cite{BLS1}, with additional
modifications in \cite{KLNS}.  By Gauss's theorem, the first term on
the right side can be written as
$$
2 \int_{\partial D} \partial_x u_0 N(x) \cdot \nabla( \partial_x u_0) dS(x)
\ ,
 $$
 where $ N(x)$ is the outward normal at the point $x \in \partial D$.
 The following computations should reveal somewhat the geometric
 structure of this term.  For any fixed point $x$ we can extend $N(x)$
 to a smooth vector field in a neighborhood of $x$.  We write
 \begin{equation} \label{starting} N(x) \cdot \nabla( \partial_x u_0) =
   \sum_{i,j} w_j N_i(x) \partial_j \partial_i u_0 = \sum_{i,j}
   w_j^\perp(x) N_i(x) \partial_j \partial_i u_0 + (w \cdot N(x))
   \sum_{i,j} N_j(x) N_i(x) \partial_j \partial_i u_0 \ ,
 \end{equation}
 where
$$
w^\perp(x) = w - (w\cdot N(x))N(x)
$$
is the projection of $w$ onto the plane tangent to $\partial D$ at the
point $x$.  For the first term on the right side of \eqref{starting}
we write
$$
\sum_{i,j} w_j^\perp(x) N_i(x) \partial_j \partial_i u_0 = -
\sum_{i,j} w_j^\perp(x) (\partial_j N_i(x)) \partial_i u_0 +
\sum_{i,j} w_j^\perp(x) \partial_j( N_i(x) \partial_i u_0)
$$
and note that the last term vanishes. Indeed, this term is a
tangential derivative of the function $N(x) \cdot \nabla u_0$ which
vanishes identically on $\partial D$ ($u_0$ is the Neumann ground
state).  Also,
$$
\sum_{i,j} w^\perp_j(x) ( \partial_j N_i(x)) \partial_iu_0 =
w^\perp(x) \cdot K(x) \nabla u_0,
$$
where $K(x) = (K_{j,i}(x)) = (\partial_j N_i(x))$ is the curvature
matrix of $\partial D$ at the point $x$.  Thus we have that
$$
N(x) \cdot \nabla( \partial_x u_0) = - w^\perp(x) \cdot K(x) \nabla u_0 +(w
\cdot N(x)) \sum_{i,j} N_j(x) N_i(x) \partial_j \partial_i u_0 \ .
$$
Once again, since $N\cdot \nabla u_0 =0$,
$$
\partial_x u_0 = w \cdot \nabla_xu_0 =w^\perp(x) \cdot \nabla_x u_0
$$
and hence
\begin{eqnarray}
  &  & 2 \int_{\partial D} \partial_x u_0 N(x) \cdot \nabla( \partial_x u_0) dS(x)  = \nonumber \\
  & -&2 \int_{\partial D} w^\perp(x) \cdot \nabla_x u_0
  w^\perp(x) \cdot K(x) \nabla u_0 dS(x) \nonumber \\
  &+& 2 \int_{\partial D} w^\perp(x) \cdot \nabla_x u_0 (w \cdot N(x))  \sum_{i,j}  N_j(x) N_i(x) \partial_j \partial_i u_0 dS(x)
  \nonumber \ .
\end{eqnarray}

\noindent To summarize we have
\begin{lemma}[\bf{Second order perturbation
    theory}] \label{perturbation2} The lowest eigenvalue as a function
  of $a$ satisfies the equation
  \begin{eqnarray} \label{eq:bigone2}
    \partial_a^2 E_0 -4 \partial_a E_0 \langle u_0, \partial_x u_0\rangle  & = & -2 \int_{\partial D} w^\perp(x) \cdot \nabla_x u_0 \,w^{\perp}(x) \cdot K(x) \nabla_x u_0\, dS(x)  \nonumber \\
    &+& 2 \int_{\partial D} w^\perp(x) \cdot \nabla_x u_0 (w \cdot N(x))  \sum_{i,j}  N_j(x) N_i(x) \partial_j \partial_i u_0 dS(x) \nonumber \\
    & & -2 \sum_{k\not= 0} \frac{B(u_k, \partial_x u_0)^2}{E_k-E_0} \ .
  \end{eqnarray}
\end{lemma}
This formula remains correct if applied to a rectangular
parallelepiped and a derivative $\partial_a$ {\it parallel} to the
edges of the domain. While the curvature matrix becomes singular along
the corners and edges of the parallelepiped one may argue that these
singularities do not contribute to the right hand side of
(\ref{eq:bigone2}) because the derivatives of $u_0$ vanish in the
directions in which $K$ is singular. A direct argument for this case
is provided in \cite{KLNS}. In this case no curvature term appears as
the faces of the parallelepiped are flat. Moreover, the second term in
\eqref{eq:bigone2} vanishes since one of the two terms $w^\perp(x)$ or
$w \cdot N(x)$ always vanishes.  Thus, one gets
\begin{corollary} \label{important} If the domain is rectangular, then
  for all derivatives $\partial_a$ parallel to the edges of the domain
  we have
  \begin{equation} \label{diffequality}
    \partial_a^2 E_0 -4 \partial_a E_0 \langle u_0, \partial_x u_0\rangle  = -  2 \sum_{k\not= 0} \frac{B(u_k, \partial_x u_0)^2}{E_k-E_0} \ .
  \end{equation}
\end{corollary}
The point of this formula is that the right side has a definite sign.
Corollary \ref{important} will be the crucial input for showing that
the energy minimizing position is when the potential sits in the
corners.

Let us expand the scope a little bit by considering smooth domains
$D$.  For such domains, by summing over the canonical basis of unit
vectors $w$ one obtains the formula
\begin{equation} \label{eq:bigone3} \Delta E_0 - 4(u_0, \nabla u_0)
  \cdot \nabla E_0 = -2 \int_{\partial D} \nabla u_0 \cdot K(x) \nabla u_0\,dS -
  2\sum_{k \not= 0} \frac{\sum_i B(u_k, \partial_i u_0)^2}{E_k-E_0} \
  .
\end{equation}
This is best seen by only using the first identity in (\ref{starting})
in the above argument without introducing $w^{\perp}$, see also
\cite{BLS1} for a direct proof.  The right side of (\ref{eq:bigone3})
has a definite sign for the case where the boundary has a positive
curvature matrix, i.e., is a convex surface.  Assuming that the right
side of (\ref{eq:bigone3}) is given, this equation can be considered
as a second order elliptic equation for the eigenvalue $E_0(a)$ and
hence it is amenable to a strong minimum principle (see, e.g.,
\cite{evans} Theorem 3 on p.~349) provided we know that the right side
of (\ref{eq:bigone3}) is strictly negative. It is then a consequence
of the strong minimum principle that if $E_0(a)$ attains its minimum
over $\overline G$ at an interior point of $G$, then $E_0(a)$ is
constant throughout $G$.

One case where one may try to exploit this reasoning is if the domain
$D$ is strictly convex in the sense that the curvature matrix $K(x)$
is positive definite at every point of the boundary $\partial D$. If
the eigenvalue $E_0(a)$ attains its minimum at $a_0 \in G$ then
$\nabla E_0(a_0) = 0$ and $\Delta E_0(a_0) \ge 0$. Hence the right
side of (\ref{eq:bigone3}) must vanish.  Since $K(x)$ is assumed to be
strictly positive at all points of $\partial D$ we have that $\nabla
u_0$ vanishes on $\partial D$, i.e., $u_0(x,a_0)$ is constant there.
Using this it is not hard to prove the following theorem.

\begin{theorem}[\bf{Strong minimum principle for $E_0$}]
  \label{thm:neumannconvex} If $E_0(a_0) = \inf_{a \in G} E_0(a)$ for
  some $a_0 \in G$, then $E_0(a)$ is identically zero. In this case
  the wave function is constant in the connected component of the
  complement of the support of the potential that touches the boundary
  $\partial D$ .
\end{theorem}

In other words, if $E_0(a)$ is not identically zero, then it attains
its minimum on the boundary of $G$.

Theorem~\ref{thm:neumannconvex} corresponds to Theorem 1.4 in
\cite{BLS1} where a proof is given. Theorem 1.4 as stated there is
slightly inaccurate since it is implicitly assumed that the complement
of the support of the potential has only one connected component (the
exterior component touching $\partial D$).

For a given domain and potential it is in general not easy to verify
if the lowest eigenvalue is independent of the position of the
potential. A construction of examples with this property, starting
with the ground state eigenfunction, was described in
Section~\ref{sec:Neumann}, see (\ref{eq:iiexample}). But it is
relatively easy to verify non-constancy of $E_0(a)$ in a number of
cases. One has to exhibit one position of the potential where the
ground state energy is not zero. This is obvious if the potential has
a fixed sign. Likewise, if the potential is not identically zero and
if its average is less than or equal to zero, we can use the constant
function as a trial function and see that the ground state energy is
strictly negative, for if it were zero, the constant function would be
the eigenfunction and the potential would be identically zero.

An interesting conclusion can be drawn using Hopf's lemma. Since the
domain is smooth, every point satisfies an interior ball
condition. Thus, by Hopf's lemma (see, e.g., \cite{evans} p. 347) we
conclude that the derivative of $E_0(a)$ at the point where the
minimum is attained and normal to the boundary is strictly positive.

While all these ideas put us on the right track, we cannot apply them
directly to our situation.  The underlying domain $\Lambda_1$ is not
strictly convex and its boundary is obviously not smooth.  Moreover,
we need the minimal configuration to be in the corners and not just on
the boundary, and most importantly we need estimates on how the
eigenvalue increases away from the boundary. Note that Hopf's lemma
will not do, since at the corners the interior ball condition is not
satisfied. Nevertheless, by a refined analysis we can show that these
statements remain true for the case where the domain is a rectangular
parallelepiped.

\subsection{$E_0$ has a non-vanishing derivatives}
Let us consider formula (\ref{diffequality}) for the case where we
take the derivative in the 1-direction. We write $G=I \times G'$ with an
open interval $I$ and an open $d-1$-dimensional rectangle $G'$. As we
shall fix the variables $a_2, \cdots, a_d$ in $G'$ we shall suppress
them from the notation and just write $E_0(a_1)$.  Thus, the variable
$a_1$ varies over the open interval $I$ which is symmetric with
respect to the origin. (\ref{diffequality}) takes the form
\begin{equation} \label{diffequality1} E_0'' -4 E_0' \langle
  u_0, \partial_1 u_0\rangle = - 2 \sum_{k\not= 0}
  \frac{B(u_k, \partial_1 u_0)^2}{E_k-E_0} \ .
\end{equation}

The following dichotomy is of interest for us.
\begin{theorem} \label{dichotomy} Assume that the right side of
  (\ref{diffequality1}) vanishes for some $a_{1,0} \in I$. Then
  $E_0(a_1) = 0$ identically in $I$ and for every $a_1 \in I$ the
  eigenfunction $u_0(x,a_1)$ is constant in the connected component of
  the complement of the support of $q_a$ that touches the boundary of
  $D$
\end{theorem}
Let us give a sketch of the proof, for more details see \cite{KLNS}.
If this function vanishes for some value of $a_{1,0}$, then we must
have that
$$
B(u_k, \partial_1 u_0) = 0, \,k=0,1,2, \dots \ .
$$
In other words
$$
\langle u_k, \Delta \partial_1 u_0 \rangle = \langle \Delta
u_k, \partial_1 u_0\rangle, \,\,k=0,1,2, \dots \ .
$$
This can be used to show that $\partial_1 u_0$ satisfies a Neumann
condition on the faces $S_1$ and $T_1$ of $G$ perpendicular to the
$1$-direction, see the proof of Lemma~2.4 in \cite{KLNS}. Thus
$u_{x_1,x_1} = 0$ on $S_1$ and $T_1$. Since $a_{1,0} \in I$ and
$(a_2,\ldots, a_d) \in G'$ the potential is zero in a neighborhood of
the boundary $\partial D$. On the faces $S_1$ and $T_1$ the function
$u_0$ satisfies the equation $-\Delta' u_0 = E_0 u_0$, where $\Delta'
= \Delta - \frac{\partial^2}{\partial x_1^2}$.  Recalling that all the
first derivatives of $u_0$ vanish on the intersections of the faces
(e.g.\ on the intersections of $S_1$ and $T_1$ with other faces of
$G$), $u_0$ must satisfy a Neumann condition on the boundary of the
faces $S_1,T_1$. Since $u_0$ is non-negative it is the ground state
eigenfunction of $-\Delta' u_0 = E_0 u_0$ and hence must be constant.
Thus, $E_0=0$ and $u_0$ is harmonic outside the support of the
potential.  If we pick a point $x_0$ on $S_1$ away from the support of
the potential (such that the potential is zero near $x_0$), we may
assume by the reflection principle, that the function $u_0(x,
a_{1,0})$ is harmonic in a neighborhood $U$ of $x_0$. Moreover,
$u_0(x, a_{1,0})$ is constant on $U \cap S_1$ and
$\partial_1 \partial_ju_0(x,a_{1,0}) = 0$ for $j=1, 2 \dots, d$. From
this one can easily conclude that $u_0(x,a_{1,0})$ must be constant in
$U$ (see \cite{KLNS}) and hence in the component of the complement of
the support of $q_a$ that touches the boundary. This proves
Theorem~\ref{dichotomy}, with the ground state being obtained by
shifting the potential and $u_0$ simultaneously.

Armed with this information we can prove Theorem \ref{thm7} easily. It
suffices to consider the $1$-direction.  Returning to
(\ref{diffequality1}) we may introduce an integrating factor $F(a_1)$
such that $F'(a_1) =-4 \langle u_0, \partial_1 u_0\rangle$ and rewrite
this equation as
$$
\left( e^{F}E'_0\right)'(a_1) = -2 e^{F(a_1)} \sum_{k\not= 0}
\frac{B(u_k, \partial_1 u_0)^2}{E_k-E_0} \ .
$$
By Theorem~\ref{dichotomy} and assumption {\bf (A1)'} the right hand
side is strictly negative.  Since we assume that the potential is
symmetric with respect to reflections about the 1-direction, we know
that the eigenvalue $E_0(a_1)$ is a symmetric function, i.e.,
$E_0(-a_1)=E_0(a_1)$ and hence $E_0'(0)=0$. Thus, for $a_1>0$, by
integrating we obtain
\begin{equation} \label{monotone} E'(a_1) = -2 e^{-F(a_1)}
  \int_0^{a_1} e^{F(\alpha)} \sum_{k\not= 0} \frac{B(u_k, \partial_1
    u_0)^2}{E_k-E_0} d\alpha <0 \ ,
\end{equation}
which implies Theorem \ref {thm7}.

\section{Open Problems} \label{sec:problems}

The main reason for presenting this expository account of our results
on the random displacement model is that the methods developed have
led to a satisfactory understanding of localization for this model,
where multiple pieces of a puzzle eventually fell into place to reveal
a complete picture. Thus, to some extend, this is the end of a
story. Nevertheless, various aspects of our work reveal natural and
non-trivial questions for further work. We end our presentation by
describing some of them.

\subsection{Bubbles tend to the boundary} \label{sec:morebubbles}

Our work has led to two types of results about the optimal placement
of a potential to minimize the lowest eigenvalue of the Neumann
problem (\ref{eq:genNeumann}) on a given domain $G$. While
Theorem~\ref{thm:neumannconvex} shows that for general convex domains
the minimizing position is at the boundary, Theorems~\ref{thm2} and
\ref{thm7} characterize corners as the exact minimizing positions for
the special case of a rectangular domain. It is natural to believe
that corners are good candidates for minimizers for other polyhedral
domains, in particular for regular $n$-gons in $d=2$, but our methods
do not allow to prove this for any $n\not= 4$. The second term in
\eqref{eq:bigone2} does not vanish unless the domain is a rectangular
parallelepiped! Of course, in this case one would assume that the
potential shares the symmetries of the domain. Radially symmetric
potentials would be the most natural candidates and it might also help
to choose them sign-definite. Particularly interesting would be a
proof of this for equilateral triangles, $n=3$, because all the other
results presented in this paper would then apply to get a localization
proof for the RDM on triangular lattices.

It is natural to conjecture that on smooth convex domains the
minimizing position of the potential along the boundary should be at a
point of maximal curvature, as long as the potential has sufficiently
small support to smoothly fit into the boundary at this point. A case
where one could hope to formulate and prove this rigorously is an
elliptic domain $G$ in $\R^2$ with a small radially symmetric
potential.

\subsection{Periodic minimizers}

Formula (\ref{eq:persupp}), characterizing the almost sure spectrum of
the RDM in terms of periodic displacement configurations, holds under
very weak assumptions on $\mu$ and the single-site potential $q$. Is
it necessary to take the closure on the right hand side? That we were
able to prove the existence of a periodic minimizer in
Theorem~\ref{thm1} was due to a number of lucky coincidences (or of
well chosen assumptions, for that matter). One might ask if for more
general cases the existence of a periodic minimizer in
(\ref{eq:persupp}) is the rule or more likely to be an exception.

For example, what happens if we keep all the assumptions in {\bf (A1)}
and {\bf (A2)} above, with the exception that supp$\,\mu$ may not
contain {\it all} the corners $\mathcal C$ of $G$? Our simple method
of constructing a minimizer by multiple reflections breaks down. It is
not clear at all, and maybe not likely, that a periodic minimizer
exists. The same problem arises if we don't require that the
single-site potential is reflection symmetric in each coordinate. One
might be led to believe that periodic minimizers hardly ever exist,
but for the moment we do not know a single counterexample.

One may also look at non-rectangular lattices, while keeping all the
desired symmetry assumptions on $q$ and supp$\,\mu$. If, as suggested
in Section~\ref{sec:morebubbles}, bubbles tent to the corners for
regular triangles, then a periodic minimizer for the triangular
lattice in $\R^2$ could be constructed by repeated reflection, as in
the case of rectangles. However, the same does not work for a
hexagonal lattice, where contradictory positions of some of the
bubbles arise after just a few reflections. Is there nevertheless a
periodic minimizer for the hexagonal RDM?

\subsection{The fractional moments method}

In our proof of localization for the RDM we followed the strategy
provided by multi-scale analysis, which yields spectral and dynamical
localization based on the Lifshitz tail bound and Wegner estimate
provided by Theorems~\ref{thm8} and \ref{thm9}. Another method which
has provided localization proofs for multi-dimensional random
operators is the fractional moments method (FMM) originally introduced
by Aizenman and Molchanov in \cite{AizenmanMolchanov} to give a simple
proof of localization for the multi-dimensional discrete Anderson
model. This method has meanwhile been extended to show localization
for continuum Anderson models \cite{AENSS, BNSS}. While likely not
being as universally applicable as MSA (and quite certainly not
extendable to situations as considered in \cite{BK}), an interesting
feature of the FMM is that, in situations where it is applicable, it
yields a stronger form of dynamical localization than what has been
obtained via MSA. Instead of (\ref{eq:dynloc}) one gets the stronger
\begin{equation} \label{eq:dynlocexp} \E \left( \sup_{|g|\le 1} \|
    \chi_x g(H_{\omega}) \chi_I(H_{\omega}) \chi_y\|_2^2 \right) \le C
  e^{-\eta |x-y|}
\end{equation}
for some $\eta>0$ and intervals $I$ in the localized regime.

It would be interesting to find out if the FMM applies to the random
displacement model under the assumptions of Theorem~\ref{thm10}. While
the FMM uses Lifshitz tail bounds in a similar way as MSA, one would
need to replace the Wegner estimate by another technical tool, the
fractional moment bound
\begin{equation} \label{eq:fmbound} \sup_{E\in I, \varepsilon >0}
  \E(\left\| \chi_x (H_{\omega}-E-i\varepsilon)^{-1} \chi_y\right\|^s)
  \le C e^{-\eta |x-y|}
\end{equation}
for suitable $s\in (0,1)$. In fact, it would already be a major step
to establish finiteness of the left hand side of (\ref{eq:fmbound}) as
an a priori bound. We believe that the properties of the RDM which
went into our proof of a Wegner estimate can also lead to a proof of
this a priori bound.

\subsection{Non-generic single site potentials}

We have argued above that alternative (i) of Theorem~\ref{thm2} is the
generic case and we have proven our localization results under this
assumption. Nevertheless, we also observed that via formula
(\ref{eq:iiexample}) one gets a rich reservoir of examples in which
the ground state of the single-site Neumann operator
$H_{\Lambda_1}^N(a)$ is constant outside the support of the potential
and thus $E_0(a)$ identically vanishing. When choosing a single-site
potential of this type in defining the random displacement model, one
sees that $\min \sigma(H_{\omega})= \min \sigma(H_{\omega,L}^N) =0$
for {\it all} choices of $\omega$ and $L$. Thus we are {\it not} in
the fluctuation boundary regime which we exploited above to get the
Lifshitz tail bound and thus can not prove low energy localization
with our methods.

The determination of the spectral type of $H_\omega$ near zero for
this case remains an open problem. The fact alone that zero is the
deterministic ground state energy for finite volume restrictions of
$H_\omega$ does not exclude the possibility of Lifshitz tails, as a
single non-degenerate eigenvalue does not affect the IDS in the
infinite volume limit. On the other hand, the ground state
eigenfunction in this model is an essentially uniformly spread out
extended state (differing from a constant only by local random
fluctuations). While we don't necessarily believe that this could
indicate the existence of absolutely continuous spectrum near zero, it
might lead to non-trivial transport, similar to the existence of
critical energies in certain one-dimensional random operators such as
the dimer model, e.g.\ \cite{JSS}.  Finding the answer to this will
require a much better understanding of excited states of the RDM, a
task which we have managed to systematically avoid in all the results
presented above.

\end{document}